\documentclass[letterpaper,twocolumn,10pt]{article}
\usepackage{atc}
\usepackage[available]{usenixbadges}

\usepackage{tikz}
\usepackage{amsmath}

\usepackage{amssymb}

\usepackage{wrapfig}
\usepackage{makecell}
\usepackage{multirow}
\usepackage{booktabs}
\usepackage{amsmath}
\usepackage{colortbl}
\usepackage{subfigure}
\usepackage{graphicx}

\usepackage{filecontents}

\usepackage{stfloats}
\usepackage{pifont}

\definecolor{mygray}{gray}{.91}

\newcommand\scalemath[2]{\scalebox{#1}{\mbox{\ensuremath{\displaystyle #2}}}}

\newcommand{\Tian}[1]{\textcolor{black}{#1}}
\newcommand{\Tam}[1]{\textcolor{black}{#1}}

\newcommand{\model}[1]{CLONE}

\newcommand{\new}[1]{\textcolor{black}{#1}}

\newcommand{\newnew}[1]{\textcolor{black}{#1}}

\begin{document}

\title{\texttt{\textbf{CLONE}}: \underline{C}ustomizing \underline{L}LMs f\underline{o}r Efficient Late\underline{n}cy-Aware Inf\underline{e}rence at the Edge}


\author{
  {\rm Chunlin\ Tian,}\hspace{2pt}  
  {\rm Xinpeng\ Qin,}\hspace{2pt}
  {\rm Kahou\ Tam,}\hspace{2pt}
  {\rm Li\ Li\textsuperscript{\dag},}\hspace{2pt}
  {\rm Zijian\ Wang,}\hspace{2pt} 
  {\rm Yuanzhe\ Zhao,}\hspace{2pt} \\
  {\rm Minglei\ Zhang,}\hspace{2pt}
  {\rm and Chengzhong\ Xu} \\[0.5em] 
  \textit{University of Macau}  
}

\maketitle 

\begin{abstract}
\Tian{
\Tam{Deploying large language models (LLMs) on edge devices is crucial for delivering fast responses and ensuring data privacy. However, the limited storage, weight, and power of edge devices make it difficult to deploy LLM-powered applications. These devices must balance latency requirements with energy consumption and model accuracy.}
In this paper, we first quantify the challenges of deploying LLMs on off-the-shelf edge devices and then we present \model~, an in-depth algorithm-hardware co-design at both the model- and system-level that intelligently integrates real-time, energy optimization while maintaining robust generality. In order to maximize the synergistic benefits of these algorithms in always-on and intermediate edge computing settings, we specialize in a 28nm scalable hardware accelerator system. 
We implement and extensively evaluate \model~ on two off-the-shelf edge platforms. Experiments show that \model~ effectively accelerates the inference process up to $11.92\times$, and saves energy up to $7.36\times$, while maintaining high-generation.}

\end{abstract}

\section{Introduction}
Large language models (LLMs) \cite{GPT4,PaLM,llama,gemma,li202512surveyreasoning} are reshaping artificial intelligence for their remarkable performance to comprehend human language and handle language-related tasks~\cite{app1-medicine, app2_clinical, app3_emergent}. Though born from the cloud, deploying LLMs on edge devices such as personal PC, smartphones, robots, and even IoT devices is becoming an important trend \cite{app4_robot,app5_robot}. First, on-device LLM can be widely used to support different kinds of applications, \new{including chatbots ~\cite{qualcomm2024llama,apple_ai}, robotics ~\cite{bilgili2024llmrobot,kabilankb2024llama}, and autonomous vehicles ~\cite{yang2023llm4drive}}. Moreover, directly accessing the LLM on the edge not only preserves data privacy but also can provide instant service without relying on a stable internet connection. 
\Tian{\new{Despite the potential benefits, deploying LLMs on commercial-off-the-shelf (COTS) edge devices faces stringent space, weight, and power (SWaP) constraints \cite{wen2024autodroid,bateni2020neuos} due to the billion-parameter size coupled with intensive computing costs.} For instance, the widely used Llama-7B model \cite{llama} requires approximately 14GB of memory for inference, even when utilizing 16-bit precision (FP16) format. However, the available RAM on typical edge devices only ranges from 4GB to 12GB~\cite{RAM}. Furthermore, inferring a single token with Llama-7B requires approximately 14 TFLOPs for an 11-token prompt~\cite{llama-dell}, which is roughly 360$\times$ the 39 GFLOPs needed by VGG-19 to process a single image with an input resolution of $224 \times 224$ \cite{FLOPs_vgg}. Moreover, LLMs' inference process demands substantial energy consumption. For instance, GPT-3 \cite{floridi2020gpt} consumes ~300 J/response \cite{ml-energy-leaderboard} on an NVIDIA A100 GPU, which is 400$\times$ the 0.75 J/response required by ResNet-50 \cite{koonce2021resnet}. Thus, intensive memory footprint, high computational demands, and significant energy costs pose severe bottlenecks to deploying LLMs at the edge.}

\noindent \textbf{Limitation of Prior Arts.}
\Tian{To break the SwaP constraints, several optimization techniques have been proposed. Techniques such as model architecture search \cite{jawahar2023llm, huang2024new, liu2024optimizing}, quantization \cite{hubara2018quantized, GPT3.int8, Just-in-time, Deja, SmoothQuant}, and pruning \cite{LLM-Pruner, AWQ, Pruner-Zero, SparseGPT} help to reduce the size of models while preserving their performance. 
However, these methods typically focus on optimizing the model itself and may not fully address the system-level trade-offs related to storage and weight.
Additionally, recent improvements in LLM compilers and software stacks \cite{pytorch-1, tensorflow, deepspeed, huggingface_transformers} have facilitated the adoption of co-processors to sustain high performance by leveraging co-processors or near-sensor processing alongside GPUs \cite{song2024powerinfer, ExeGPT, PagedAttention, Just-in-time, llm_flash, Splitwise, PagedAttention, flexgen, Orca,zhao1,zhao2,zhao3}.
However, they incur additional computational/communication overhead that significantly reduces the edge device lifecycle and interferes with the smooth execution of other applications. To address power constraints, Dynamic Voltage and Frequency Scaling (DVFS) \cite{dvfsasplos, dfvs-4, dvfs-2, dvfs-3, bateni2020neuos} has been widely adopted to adjust the power usage of processors by dynamically changing their voltage and frequency.
However, most existing approaches are designed for discriminative CNN or RNN and treat the entire network as a black box during tuning. Generative LLMs have not undergone the necessary scrutiny due to the auto-regressive inference schemes and the heterogeneity of stochastic requests and generated outputs. \textit{Thus, an edge-tailored LLM system that intelligently coordinates model- and system-level optimizations, balancing energy efficiency, latency, and model performance is urgently required.}}

\noindent\textbf{Challenge.} Designing such a customization system is not straightforward and faces the following critical challenges. First, the “billion-parameter” poses a significant barrier to deployment on resource-constrained edge devices due to the high memory footprint and computation cost. How to customize the LLM according to the hardware profile of a specific edge device to effectively meet the memory constraint and reduce the computing latency while ensuring the generating capability is the first critical challenge. 
\Tian{In addition, when deploying customized LLMs on edge devices, LLM-powered applications feature stochastic and dynamic I/O, unlike traditional models with stable inputs. Moreover, tailored layers introduce unique approximation characteristics, while interference from co-running applications during mobile execution can cause runtime variance. Therefore, designing an efficient system-level controller to meet latency targets while optimizing energy consumption is another challenge. Furthermore, model-level (accuracy/latency) and system-level (energy/latency) optimizations in isolation can be naive,  often resulting in unnecessary energy consumption or reduced accuracy. Effectively coordinating these optimizations to balance accuracy, latency, and energy efficiency is the third challenge.}

\Tian{In this work, we present \model~, a comprehensive system that intelligently \underline{c}ustomizes \underline{L}LM f\underline{o}r efficient late\underline{n}cy-aware inf\underline{e}rence on commercial-off-the-shelf edge devices. It adopts a hierarchical structure to jointly handle model- and system-level optimization in a unified manner to consider real-timeliness, model accuracy, and energy efficiency jointly. 
Specifically, \model~ consists of two main phases: offline device-specific model tailoring and online latency-aware system optimization.
Within the offline tailoring process, for a specific device, \model~ reformulates the tailoring of target LLMs for a specific device as a generative task within a continuous representation space. Using an encoder-evaluator-decoder architecture, it generates optimal pruning configurations through gradient-based optimization, effectively bridging the gap between resource constraints and model performance. To support diverse edge applications\new{~\cite{apple_ai,wen2024autodroid,yi2024phonelm}}, \model~ performs parameter-efficient fine-tuning on the customized LLM using multiple plug-and-play Low-Rank Approximation (LoRA) \cite{LoRA} adapters, ensuring adaptability and optimized performance.
Once the customized model is ported to edge devices, during online inference, \model~ employs a Mixture-of-Experts (MoE) \cite{moe,tian2024hydralora} router to dynamically integrate optimal LoRA adapters, enabling more precise responses to stochastic, complex, or mixed-task end-user requests.
At the system level, \model~ applies Dynamic Voltage and Frequency Scaling (DVFS) to the per-token, autoregressive inference process of LLMs, optimizing the supply voltage ($V_{DD}$) and operating frequency ($F_{req}$) to minimize energy consumption while meeting real-time constraints. Recognizing the LLM layer characteristics, especially post-pruned uneven parameter, \model~ enhances the granularity of DVFS adjustments at layer boundaries. This provides greater flexibility and adaptability for power-sensitive systems compared to traditional workload-level (whole model) black-box optimizations. While existing vanilla DVFS on edge devices are typically discrete, \model~ introduces a learning-based DVFS approach to reduce budget gaps that maximize system efficiency \newnew{and} model performance. 
To fully harness the benefits of these technologies, we have developed a specialized 28nm scalable hardware accelerator system. It contains a LoRA Processing Unit (LPU) for hot-swapping adapters, enabling dynamic model performance adjustments through dedicated data paths, and a Special Function Unit (SFU) designed for fine-grained and continuous DVFS adjustments equipped with a fast-switching low-dropout (LDO) voltage regulator and an all-digital phase-locked loop (ADPLL). The LPU supports the runtime MoE router algorithm to handle stochastic requests, eliminating bottlenecks of general-purpose processors and boosting computational efficiency. The SFU implements the learning-based DVFS algorithm to effectively manage runtime variance and swiftly achieve the desired $V_{DD}$ and $F_{req}$.} Specifically, we make the following key contributions:
\Tian{
\begin{itemize}
    \item We propose \model~, an efficient software-hardware co-design system for customized LLM inference on resource-constrained edge devices, where the model- and system-level coordinators cooperate to intelligently balance the inference speed and energy efficiency while maintaining robust generative performance.
    \item We design hierarchical offline/online optimization phases that effectively coordinate customization by addressing static model weights, hardware characteristics, stochastic I/O patterns, and runtime variance. Additionally, a 28nm accelerator with specialized datapaths supports these algorithms for optimized performance.
    \item To evaluate the effectiveness of \model~, we conduct extensive experiments based on commercial-off-the-shelf edge devices, representative LLMs and benchmarks. 
\end{itemize}}

\section{Background and Related Work}

\subsection{Large Language Models}
\label{section:llm}
\noindent\textbf{LLM Architecture.} \Tian{In contrast to traditional deep neural networks (DNNs) and convolutional neural networks (CNNs)~\cite{lecun2015deep,goodfellow2016deep}, which integrate diverse types of layers (e.g., convolutional (CONV), fully connected (FC), recurrent (RC), pooling, etc.) designed for specific tasks, large language models (LLMs) predominantly consist of a uniform stack of transformer decoder layers. For instance, Llama-7B \cite{llama} adopts a homogeneous architecture composed of 32 identical LlamaDecoderLayers. Each decoder layer encompasses two core components: LlamaAttention and LlamaMLP. Despite the structural uniformity across decoder layers, their contributions to model efficiency and effectiveness vary significantly ~\cite{GPT3.int8,Outlier}. Consequently, optimizing the inference execution of LLMs necessitates a detailed analysis of the individual impact of each layer ($\S \ref{Section:layer}$).}

\begin{figure}[!t]
    \centering
\includegraphics[width=0.95\linewidth]{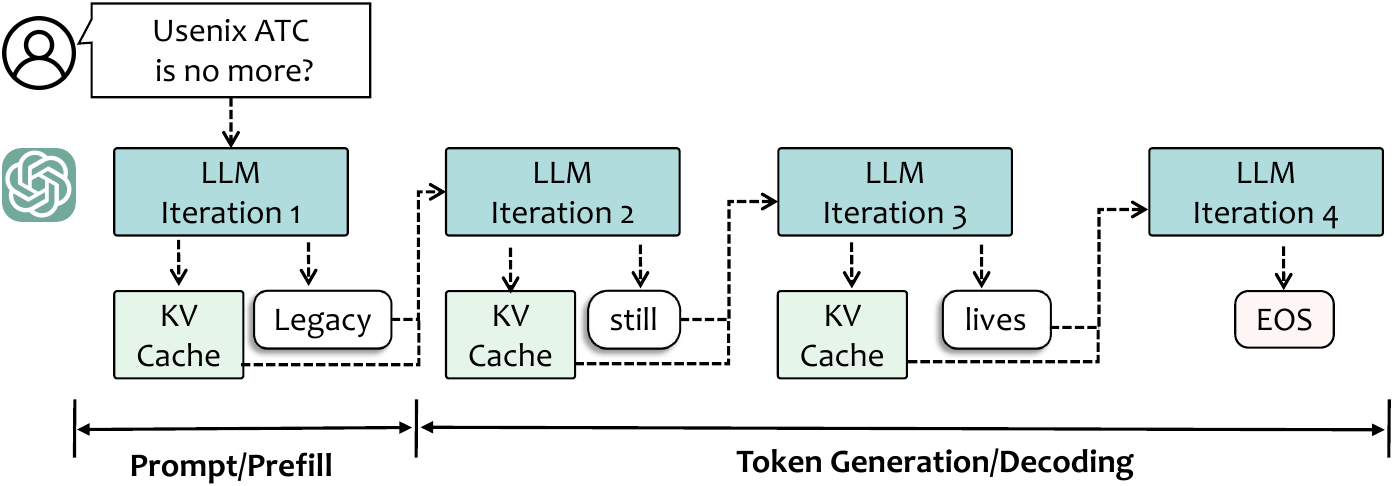}
    \caption{\new{Overview of LLMs autoregressive inference.}}
    \vspace{-1.5em}
    \label{fig:inference}
\end{figure}

\noindent \textbf{LLMs Inference.}
\label{section:inference} 
LLMs process a structured sequence involving multiple forward passes through the model to sequentially generate each output token. Figure \ref{fig:inference} shows the inference process with a simple example. Typically, this process mainly contains two stages~\cite{flexgen,sosp,alammar2019illustrated}.
\textbf{1) Pre-fill} takes a prompt sequence and generates the key-value (KV) cache for each Transformer layer of LLM. Upon receiving the prompt \new{\textit{``Usenix ATC
 is no more?''}}, the tokenizer embeds input as tokens, denoted as $X_{in} \in \mathbb{R}^{n\times d}$, where $d$ is the hidden size and $n$ is the length of input token. Then, the LLM handles all input tokens in parallel during a single forward iteration to generate a KV cache.
The output of attention is sent to MLP to generate the first output token \new{\textit{``Legacy''}}. Large-scale matrix multiplications are required to generate the KV cache, which makes the pre-fill computing intensive. 
\textbf{2) Decoding} utilizes and updates the KV cache to generate tokens step-by-step.
Following the generation of the first token, the LLM leverages the KV caches prepared earlier and adds new information to them. The creation of each new token is influenced by the tokens generated before it. During each token generation, for the input $X_{dec} \in \mathbb{R}^{1 \times d}$, attention layers load the previously stored KV cache, and new KV pairs are computed and concatenated to the existing cache. The output of the last decoder layer is sent to the final prediction layer to predict the next token sequentially. It executes iteratively until an End of Sequence (EOS) token is encountered or a predefined termination criterion is met. \Tian{Unlike traditional models with fixed input formats and structured workflows~\cite{lecun2015deep,goodfellow2016deep}, LLM inputs and outputs are highly non-deterministic. This stems from the diverse, open-ended nature of user prompts, which vary widely in structure, intent, and context~\cite{prompt_formats,input_unpredict,P-Tuning}. Additionally, autoregressive token generation is inherently probabilistic, driven by sampling methods ~\cite{llama,Splitwise,output_token,output_sampling} and variability in training datasets~\cite{attention,output_train_1,output_train_2}, making outputs context-sensitive.}

\subsection{Bottlenecks of Deploying Edge LLMs}

Table \ref{tab:device} lists specifications for server-level and edge-level processors commonly used for ML workloads, highlighting resource collapse. Despite the potential benefits, \newnew{including privacy preservation and instant responses without depending on a stable internet connection}~\cite{Twig,Mistify}, deploying LLMs on the edge faces the following critical bottlenecks.


\noindent \textbf{1) High Memory Footprint.} 
The main contributors of ``billion-parameter'' LLMs are model weights (memory is occupied by the model parameters) and KV cache (memory is occupied by the caching of self-attention tensors to avoid redundant computation). 
For example, Llama-7B in 16-bit precision requires approximately $14$GB memory ($7B \times \text{sizeof(FP16)}$). Its architecture with 32 layers, 32 heads per layer, and a head dimension of 128 incurs a memory cost of $0.5$MB per token, accounting for K and V matrices. Consequently, processing 4096 tokens demands 2GB, limiting the size of models that can be deployed on edge devices with 4–12GB memory \cite{RAM} and often causing Out-of-Memory (OOM) errors.

\begin{table}[!ht]
    \centering
    \caption{Popular ML hardware specifications.}
    \resizebox{0.95\linewidth}{!}{
    \begin{tabular}{c|cccc}
    \bottomrule[1.5pt]
    \rowcolor{mygray}  \textbf{GPU Types}   &  \textbf{Peak Perf.} & \textbf{Memory} & \textbf{Bandwidth} & \textbf{Peak Power}\\
    \midrule[0.75pt]
    \multicolumn{5}{c}{Server-level} \\ \hline
    NVIDIA  A100   & 312 TFLOPS &80GB & 1935 GB/s & 300W\\
    NVIDIA  A40   & 149.7 TFLOPS & 48 GB  & 696 GB/s & 300W\\ \hline
    \multicolumn{5}{c}{Edge-level} \\ \hline
    Jetson Orin NX & 100 TOPS & 16GB & 102.4GB/s&25W\\
    Jetson Orin Nano & 40 TOPS &8GB &68 GB/s &  15W\\
    \toprule[1.5pt]
    \end{tabular}}
    \label{tab:device}
\end{table}

\begin{figure}[!t]
    \centering
    \subfigure[Server/edge latency.]{\includegraphics[width=0.4\linewidth]{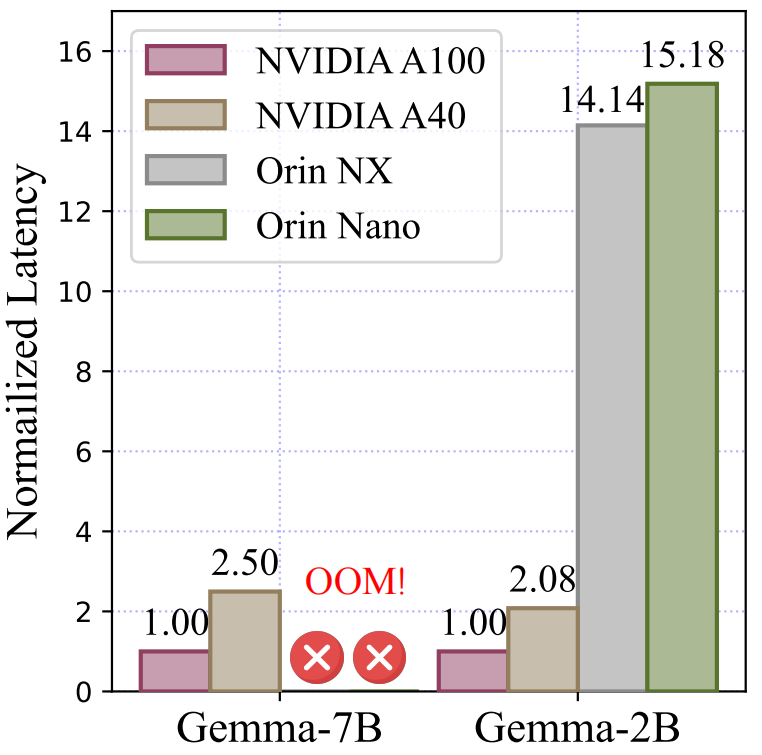}}
    \subfigure[\new{Latency breakdown.}]{\includegraphics[width=0.56\linewidth]{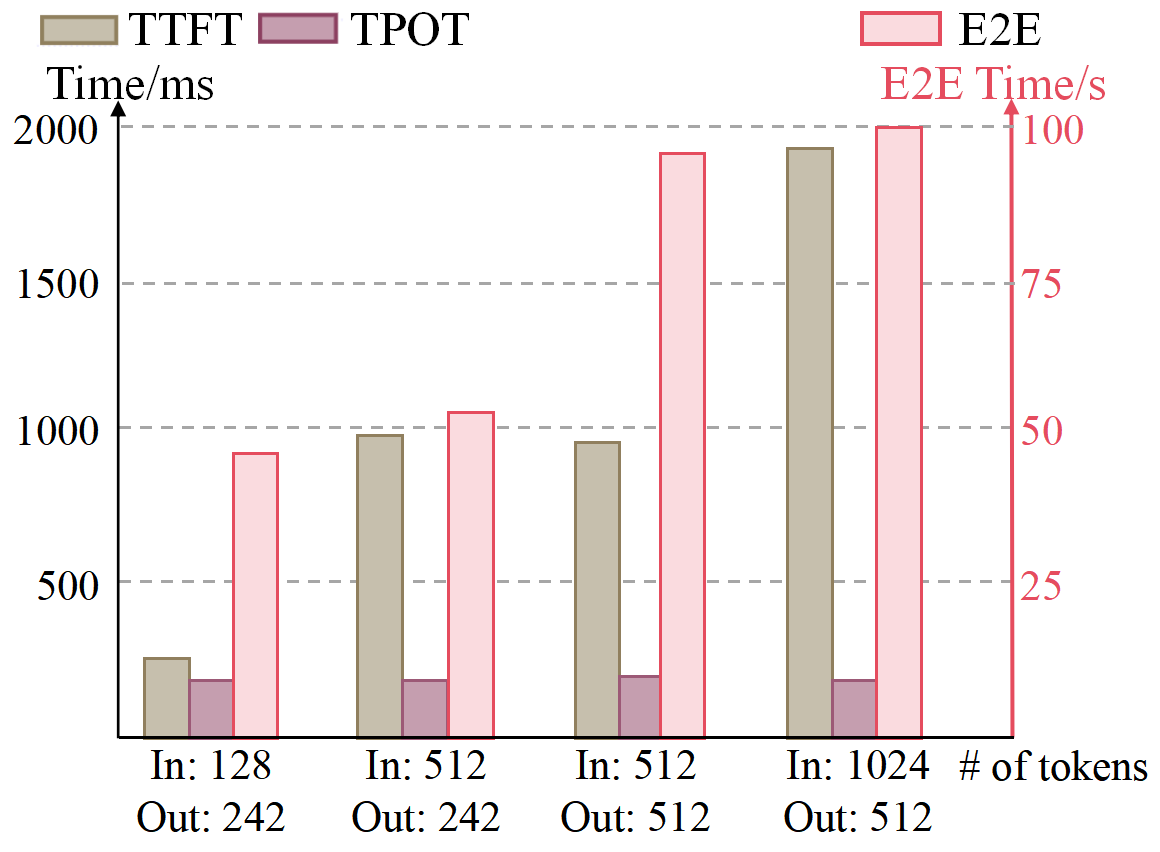}}
    \caption{\textcolor{black}{LLMs inference latency analysis. (a) Latency efficiency of two common LLMs (Gemma-2B/7B) inference use cases over server-level and edge-level processors. (b) Inference latency with different prompt and output tokens.}}
    \label{fig:edge_TE}
\end{figure}

\noindent \textbf{2) High Inference Latency.} Inference latency is a crucial metric for mobile optimization because if the latency of a service exceeds the human-acceptable thresholds (e.g., 33.3 ms for a 30 FPS video frame rate \cite{User-aware,Event-based} or 50 ms for interactive applications \cite{OSDI_96,micro_15}), end-users will abandon the service \cite{Event-based}. To evaluate inference QoE for LLMs, there are three main metrics: 1) \textit{Time To First Token (TTFT)} latency from input to the output of the first token (prefill). Low TTFT means fast response, which is essential for user experience in real-time interactions. 2) \textit{Time Per Output Token (TPOT)} latency to track the auto-regressive process of each token generated serially. TPOT refers to how each user perceives the ``speed'' of the model. 3) \textit{End-to-end (E2E)} latency, the overall time it takes for the model to generate the full response for a user, denoted as: $E2E =(TTFT) + (TPOT) * N,$
where $N$ is the generated token number. Figure \ref{fig:edge_TE} (a) illustrates the latency across different processors running the same LLMs, representative Gemma-2B/7B \cite{gemma}, on the Wikitext2 \cite{Wikitext2} dataset. While Gemma-7B is OOM for edge device. For the same user request on Gemma-2B, the Orin Nano experiences 15.18$\times$ more latency than the NVIDIA A100, significantly impacting the Quality of end-user Experience (QoE) \cite{autoscale,energy_survey,CAMA,eNVM}. Additionally, LLMs operate with an auto-regressive and probabilistic inference process~\cite{song2023llm}. Figure \ref{fig:edge_TE} (b) shows as prompt sizes increase (128 $\rightarrow$ 512 $\rightarrow$ 1024) inference Gemma-2B on Orin NX. We can observe that the number of pre-filled tokens increases significantly (255$\rightarrow$1935ms) due to parallel processing capabilities. Conversely, the TPOT remains relatively stable (180$\sim$200ms) due to serial decoding generation. However, the E2E is more than 100s, which fails to serve the end-user demands, indicating opportunities for optimizing practical inference performance.

\noindent \textbf{3) High Energy Consumption. } 
In server-level GPU deployments, it is typically assumed that these units are plugged into wall power due to the substantial energy demands of LLMs inference\cite{ExeGPT, Splitwise, PagedAttention, Micro_LLM, llm_flash, wu2025survey}. However, this assumption does not hold for edge-level devices such as robots and autonomous vehicles, where energy availability is severely restricted \cite{edge_energy}. In contrast to conventional on-device tasks, LLM inference consumes an order of magnitude more energy. For example, a Google search driven by a large AI model expends 8.9 watt-hours (Wh) of energy, approximately 30 times the 0.3 Wh required by a standard Google search \cite{AIIndex2024}. Importantly, ``Scaling Laws'' \cite{Scaling,Scaling_up} suggest that as model parameters scale up, there is a corresponding increase in both performance and energy consumption. Transitioning from Gemma-2B to 7B, for instance, boosts accuracy from 42.3\% to 64.3\% on the MMLU benchmark \cite{MMLU}—a comprehensive evaluation platform—at the cost of tripling energy usage. Consequently, there is a pressing need to optimize the energy efficiency of LLM inference at edge to meet QoE.

\section{Motivation}
\label{section:opportunity}
In this section, we present key observations from model-intrinsic dimensions (layer characteristics ($\S \ref{Section:layer}$)), stochastic I/O ($\S \ref{Section:IO}$)) and system factors (hardware and runtime variance ($\S \ref{Section:runtime}$)) for practical LLM inference on edge devices. The design space is analyzed across three critical axes: generation quality, latency, and energy efficiency.

\subsection{Heterogeneity of Model Characteristics}

\label{Section:layer}

\begin{figure*}[!ht]
    \centering
    \begin{minipage}{0.64\linewidth}
        \centering
        \subfigure[Generative-ability]{\includegraphics[width=0.323\linewidth]{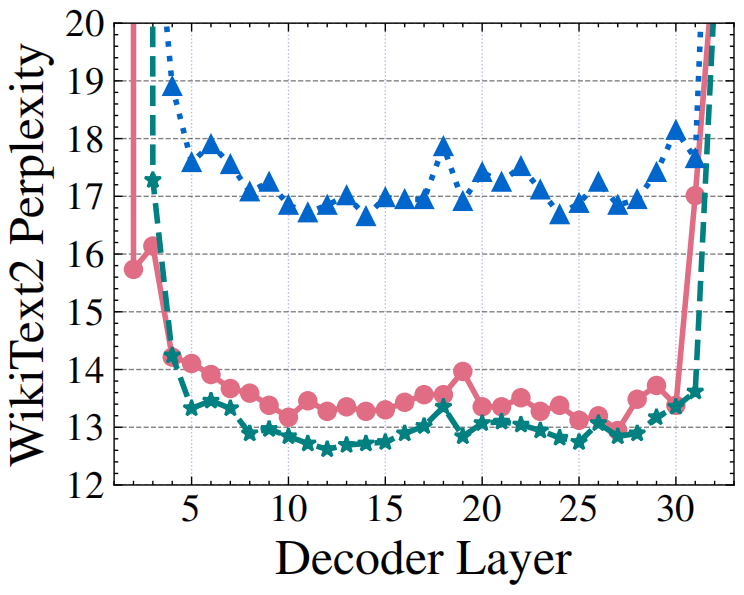}}
        \subfigure[Energy]{\includegraphics[width=0.335\linewidth]{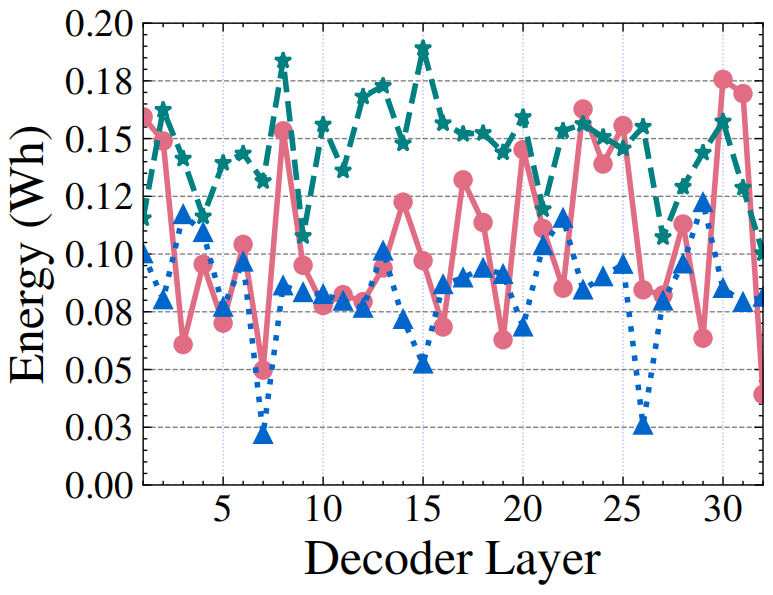}}
        \subfigure[\textcolor{black}{Latency}]{\includegraphics[width=0.325\linewidth]{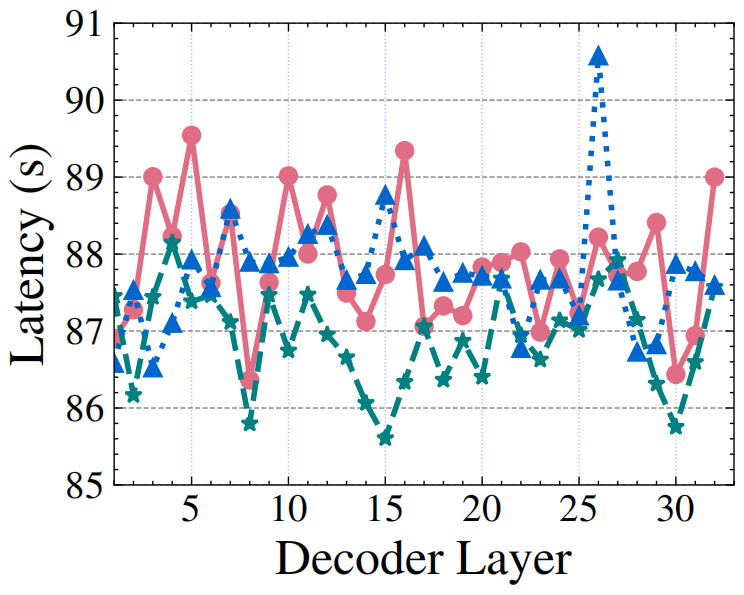}}
        \caption{\new{Layer sensitivity analysis: removing specific layers one-by-one to measure each layer's impact on Llama-7B (\textcolor{red}{$\bullet$}), Llama2-7B (\textcolor{teal}{$\star$}), and Vicuna-7B (\textcolor{blue}{$\blacktriangle$}).}}
        \label{fig:layer_break}
    \end{minipage}%
    \hspace{0.04\linewidth}
    \begin{minipage}{0.25\linewidth}
    \centering
    \includegraphics[width=0.75\linewidth]{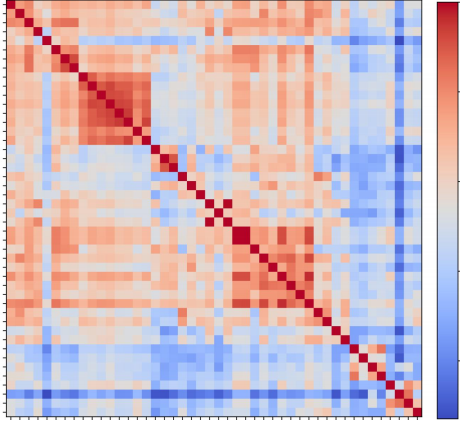}
    \caption{Task embedding similarity heatmap.}
    \label{fig:lora_embedding}
    \end{minipage}
\end{figure*}

In order to investigate the contribution of different decoder layers, we conduct an in-depth analysis of the inherent sensitivity of the stacked transformer architecture across three critical dimensions: generative ability, energy efficiency, and latency. 
For general-purpose, we employ zero-shot perplexity (PPL)~\cite{ppl-1,llama,gemma}, a common metric, to evaluate generative abilities, lower PPL indicates higher model adeptness in predicting sequence tokens. Figure~\ref{fig:layer_break} (a) presents the PPL on the WikiText2~\cite{Wikitext2} dataset after removing specific decoder layers in the Llama-7B, Llama2-7B~\cite{Llama_2}, and Vicuna-7B~\cite{vicuna} models. The results show that the front and back layers have higher PPL values than the middle layers. Because the front layers play a critical role in feature extraction, while the back layers significantly influence output generation, both contribute markedly to model performance. while LLMs maintain a homogeneous layer structure, the generative impact varies by layer due to input and output data stream heterogeneity.
To evaluate the system effectiveness, we infer LLMs on WikiText2, monitoring energy consumption and duration using CodeCarbon~\cite{codecarbon}. Figure~\ref{fig:layer_break} (b, c) shows that removing different decoder layers yields non‑uniform changes in end‑to‑end energy and latency. This suggests that, beyond nominal parameter counts, each layer contributes heterogeneously to practical resource usage. \new{We attribute this to (i) varying sequence‑length–dependent workloads, (ii) hardware‑level cache and parallel‑sync effects, and (iii) non‑linear inter‑layer interactions typical of deep networks.}

\emph{$\blacktriangleright$ \new{Motivation} 1}: \Tian{LLM layers contribute unevenly to effectiveness and efficiency, underscoring opportunities for model customization by pruning non-essential components and system optimization through fine-grained layer-wise tuning.}

\subsection{Heterogeneity of Stochastic Input/Output}
\label{Section:IO}
\begin{figure*}[!ht]
    \centering
    \subfigure[Practical inference traces.]{
        \includegraphics[width=0.22\linewidth]{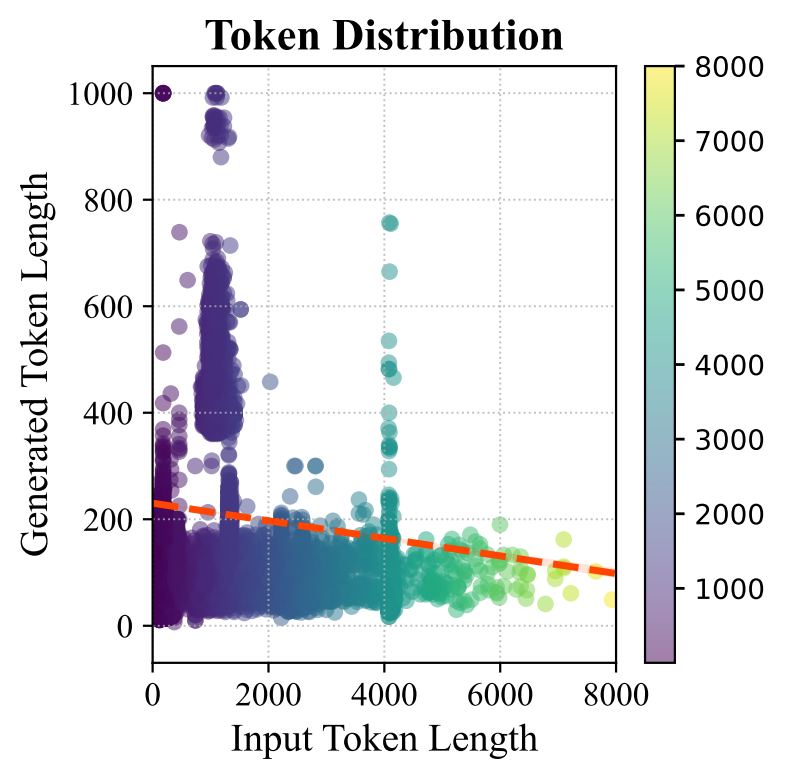}
    }
    \subfigure[Memory consumption.]{
        \includegraphics[width=0.22\linewidth]{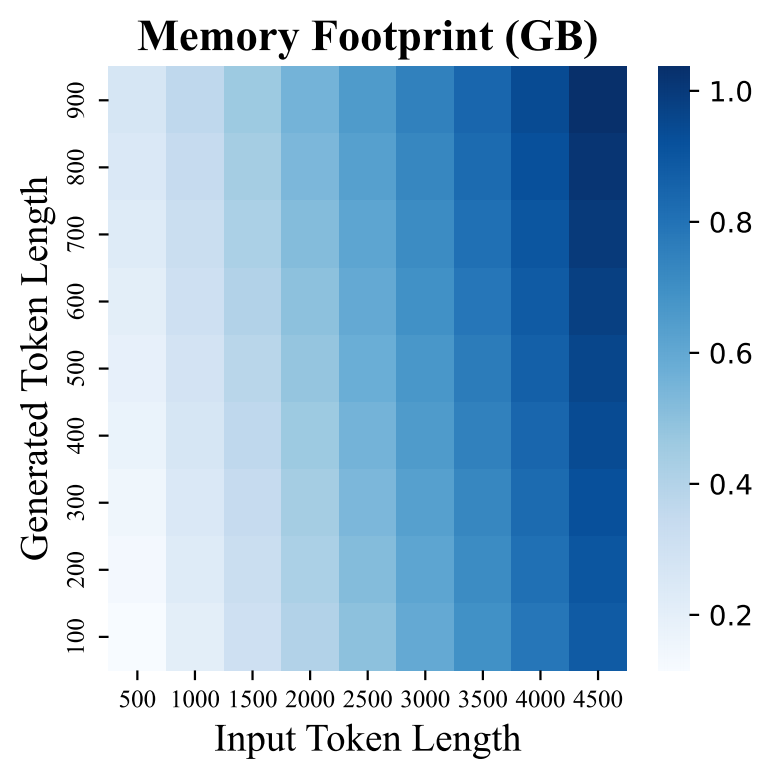}
    }
    \subfigure[Latency analysis.]{
        \includegraphics[width=0.22\linewidth]{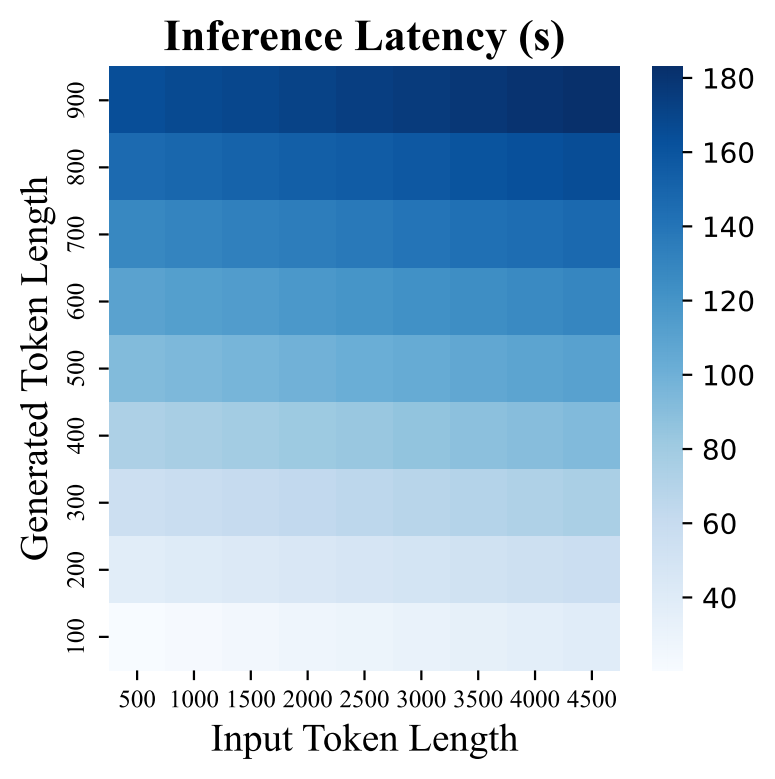}
    }
    \subfigure[Energy consumption.]{
        \includegraphics[width=0.22\linewidth]{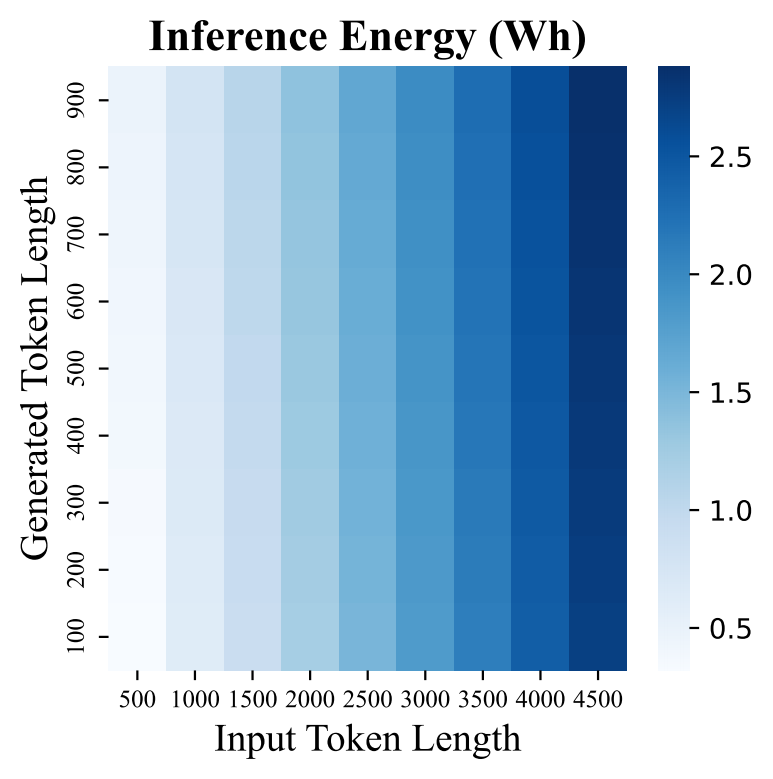}
    }
    \caption{Comprehensive analysis of LLM inference system using Gemma-2B on NVIDIA Orin NX. (a) Analysis of practical LLM inference traces., while (b-d) highlight memory, latency, and energy consumption w.r.t. input and generated length.} 
    \label{fig:ob_sys_analysis}
\end{figure*}

\Tian{As discussed at $\S \ref{section:inference}$, inference serving systems exhibit significant non-deterministic due to the diversity of input prompts, and LLM auto-regressive output exhibit distinct execution behaviors.
\noindent \textbf{Case study.} As shown in Figure \ref{fig:ob_sys_analysis} (a), we first illustrate the distribution of prompt input and generated output tokens on an Azure LLM inference services trace \cite{Splitwise}. These long-tail patterns and the unpredictability of token generation lengths highlight the need for dynamic resource allocation to manage varying computational demands. 
\noindent \textbf{Performance impact.} To analyze the inherent patterns of mixed-task end-user requests, we visualized the task embedding similarities of \new{the Flanv2 dataset~\cite{flanv2} using a heatmap, as shown in Figure \ref{fig:lora_embedding}. Each axis enumerates the individual subtasks within the dataset, and each matrix cell quantifies the pairwise correlation between subtasks.} The results reveal significant data heterogeneity across tasks.
\noindent \textbf{System impact.} Otherwise, requests of different input and output lengths possess different compute and energy characteristics. As shown in Figure \ref{fig:ob_sys_analysis} (b-d), we measure Gemma-2B \cite{gemma} on Nvidia Orin NX with different input and generated length to evaluate the memory footprint, latency and energy consumption. Memory consumption is influenced by both static model parameters and the dynamic KV cache, which grows as the number of tokens to be processed increases. E2E latency is primarily impacted during the decoding phase, as it is step-by-step, whereas the prefill phase computes all prompts in parallel. Consequently, an increase in generated tokens significantly impacts overall latency. Energy consumption is higher during the prefill phase due to large-scale matrix operations, while the decoding phase, processing one token at a time, has lower power demands. Inefficient system performance degrades QoE, making it critical to meet diverse LLM request demands without compromising output quality.}

\emph{$\blacktriangleright$ \new{Motivation} 2}: \Tian{Stochastic I/O with mixed-task requests underscores the need for precise inference services to improve model performance. Similarly, significant resource slack suggests potential for fine-grained, LLM-specific system optimizations to improve utilization efficiency.}

\subsection{Heterogeneity of Runtime Platform}
\label{Section:runtime}
\begin{figure*}[!t]
    \centering
    \begin{minipage}{0.35\linewidth}
        \centering
    \includegraphics[width=1.0\linewidth]{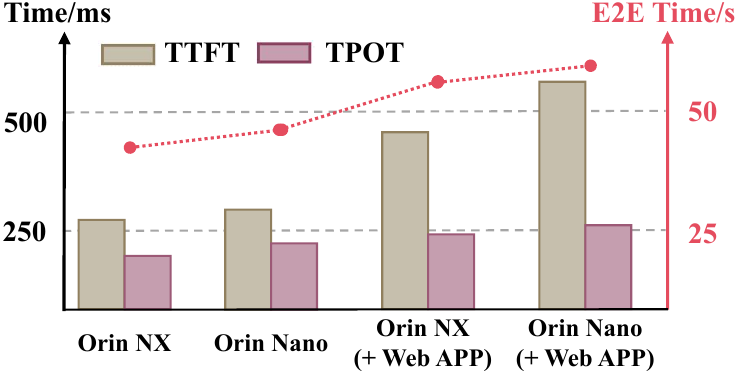}
        \caption{\textcolor{black}{\new{Impact of different hardware devices (Nano/NX) and concurrently running apps (web search) on edge LLMs inference.}}}
        \label{fig:heter}
    \end{minipage}%
    \hfill
    \begin{minipage}{0.62\linewidth}
        \centering
            \centering
    \includegraphics[width=0.32\linewidth]{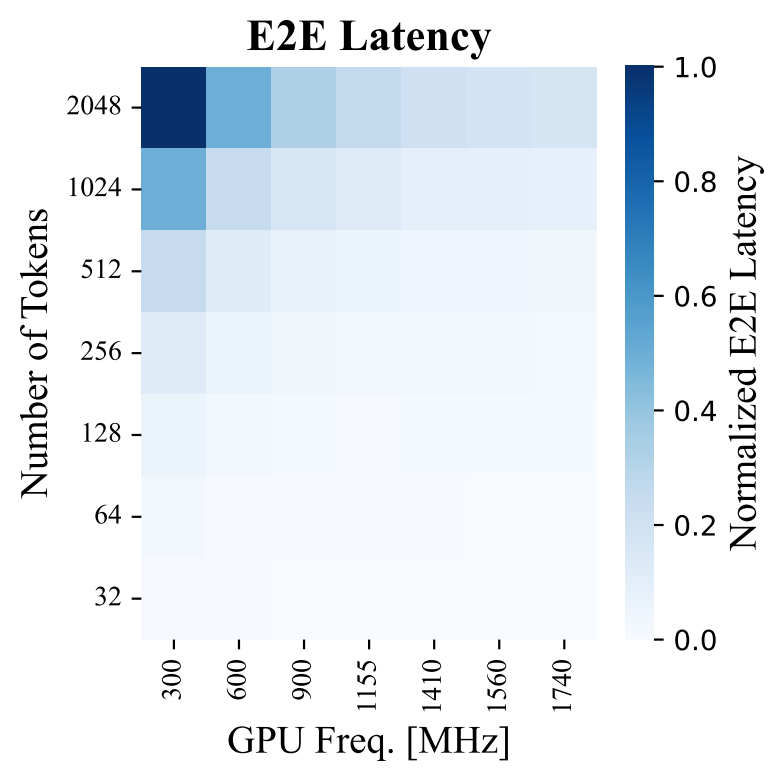}
    \includegraphics[width=0.32\linewidth]{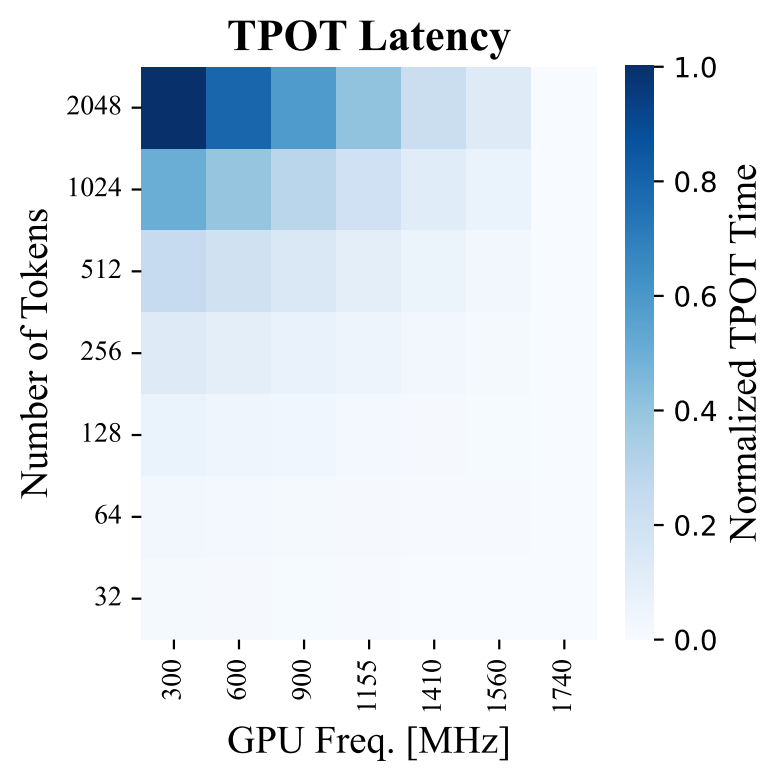}
    \includegraphics[width=0.32\linewidth]{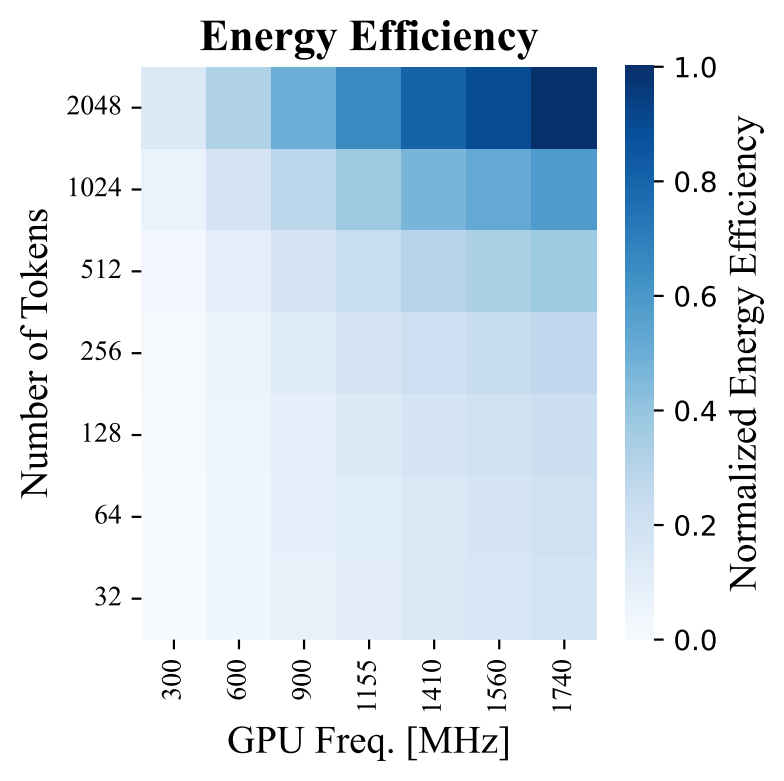}
    \caption{Impact of different token lengths and GPU frequencies on a) end-to-end (E2E) latency, b) time per output token (TPOT), and c) energy efficiency.}
    \label{fig:freq}
    \end{minipage}
\end{figure*}

\Tian{Practical system heterogeneity encompasses performance and energy variations arising from platform diversity (static heterogeneity, Table \ref{tab:device}), diverse LLM application demands, and interference from co-running applications (dynamic heterogeneity).
Table~\ref{tab:app} lists LLM requests from various applications or tasks, each with distinct QoE demands \cite{ELMS,application_1,Event-based,application_3,tian2022harmony,wu2024heterogeneity,yeboSurvey,zhan2024heterogeneity}.}
To investigate the impact of edge device hardware and concurrent running apps, \new{we
deployed the Gemma‑2B model on two NVIDIA platforms—Jetson Orin NX and Orin Nano—and processed a workload comprising 128 prompt tokens and 242 generated tokens (the first scenario in Figure~\ref{fig:edge_TE}(b)). We profiled the resulting inference behavior, and Figure~\ref{fig:heter} reports the TTFT, TPOT, and E2E inference latency for both devices, alongside the corresponding performance degradation observed in a foreground web‑search application~\cite{Event-based,web_app_2}.}
We find that 1) Due to the varying processing capabilities of different edge devices, the response time to the same end-user request can differ significantly. For instance, when both are in an idle state, the Orin NX, with its superior computational power, provides more timely feedback (TTFT: 255; TPOT: 198 ms) compared to the Orin Nano (TTFT: 268ms; TPOT: 231ms). 2) Resource contention caused by user interaction with the concurrent running app prominently slows down the model inference progress. For instance, the E2E inference latency on Orin NX is 42.6s when there is no user interaction in the foreground. However, it increases to 61.4s with a website searching application running in the foreground. This highlights the impact of latency on user experience, as service abandonment occurs when latency exceeds the human-acceptable threshold \cite{Event-based,application_3}.
\Tian{Figure \ref{fig:freq} demonstrates that higher GPU frequencies reduce E2E latency and TPOT by accelerating token processing (a, b) while supporting energy-efficient control via frequency scaling (c). Dynamically adjusting processor frequencies based on workload size enhances efficiency with minimal performance impact, where the optimal frequency is determined by the LLM architecture and output length. Although DVFS~\cite{dfvs-4, dvfs-2, dvfs-3} is commonly used to scale voltage and frequency for low-intensity workloads, existing schemes often rely on coarse-grained, black-box, workload-level adjustments with heuristic methods~\cite{dfvs-4, dvfs-2, dvfs-3, bateni2020neuos, dvfsasplos}.}

$\blacktriangleright$ \textit{\textbf{\new{Motivation} 3:}} \Tian{Hardware type and LLM application SLOs significantly impact end-user QoE, while resource preemption from foreground applications exacerbates runtime system heterogeneity. Accurate runtime estimation of application demands and device capabilities is crucial for guiding DVFS and optimizing LLM inference system efficiency.}

\begin{table}[!t]
    \centering
    \caption{SLOs of various LLM applications. Time To First Token (TTFT): refers to the latency from input to the first token output (prefill), while Time Per Output Token (TPOT) is the latency for each subsequent token (decode).}
    \resizebox{0.95\linewidth}{!}{\begin{tabular}{c|c}
    \bottomrule[1.5pt]
    \rowcolor{mygray} \textbf{LLM Application}& \textbf{Service-Level Objective}\\
    \toprule[0.75pt]
        Chatbot \cite{chatgpt, anthropic_claude} &  Readable TTFT/TPOT\\
        Search Engine \cite{microsoft_bing_new_features} & Low TTFT, Medium TPOT \\
        Event Logger \cite{otter2024,rewind2024,mem2024} & Tolerable TTFT, Medium TPOT \\
        Smart  Reply~\cite{googleMLKit2024,azureCognitive2024} & Low TTFT, Low TPOT\\
        Code Generator \cite{github_copilot} & Medium TTFT, High TPOT \\
        Virtual Assistant~\cite{googleAssistant2024,siri2024} & Low TTFT, Medium TPOT\\
    \toprule[1.5pt]
    \end{tabular}}

    \label{tab:app}
\end{table}

\section{\model~: Design}
\subsection{System Overview}
\begin{figure}[!t]
    \centering
    \includegraphics[width = 0.95\linewidth]{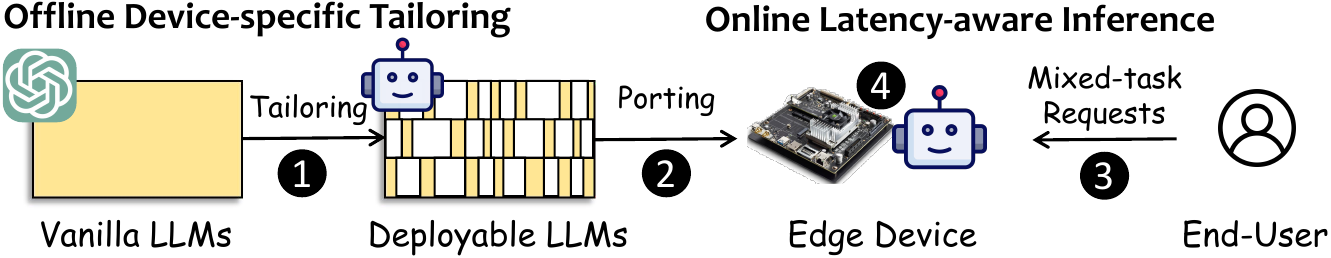}
    \caption{The overview of \model~, including offline device-specific tailoring and online latency-aware inference.}
    \label{fig:overview}
\end{figure}

Figure~\ref{fig:overview} presents the architecture of \model~ which can be mainly divided into the following two phases: 1) offline device-specific tailor and 2) online latency-aware optimization. 
The overall workflow of \model~ can be represented as the following main steps. \Tian{\ding{202} \model~ leverages edge hardware profiles and LLM architecture using an encoder-evaluator-decoder framework to reframe LLM compression as a data-driven generative task. This approach addresses memory constraints, meets latency and energy requirements, and preserves generation performance. Plug-and-play LoRA adapters further fine-tune the tailored LLM for diverse edge applications.
\ding{203} The deployable model is then ported onto the edge device.
\ding{204} During runtime inference, to handle stochastic and mixed-task user requests, \model~ integrates a MoE router to dynamically select optimal LoRA adapters, enhancing model performance.
\ding{205} At the system level, \model~ incorporates a learning-based DVFS controller to adjust device voltage ($V_{DD}$) and frequency ($Freq$) dynamically at layer boundaries during per-token autoregressive inference. This minimizes energy consumption while ensuring target latency is met.}
Furthermore, in order to fully leverage the synergistic advantages of these algorithms, we develop a specialized 28nm scalable hardware accelerator system designed to further boost 
energy efficiency through customized hardware integration.

\subsection{Offline Device-specific Tailoring}

\begin{figure*}[ht]
    \centering
    \includegraphics[width = 0.85\linewidth]{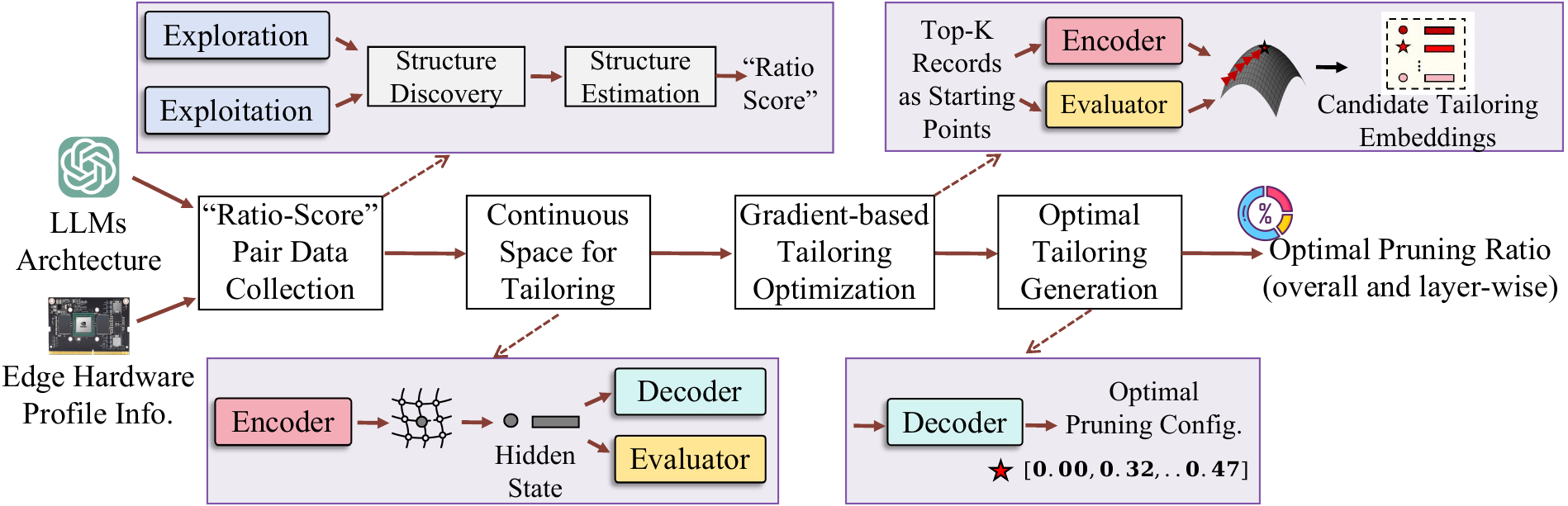}
    \caption{Hardware-aware tailor workflow. 
    Four key components: ``ratio-score'' data collection, continuous space for tailoring,  gradient-based tailoring optimization, and optimal tailoring generation.}
    \label{fig:offline}
\end{figure*}

According to \textit{\newnew{Motivation} 1}, LLMs demonstrate considerable parameter redundancy, especially within their intermediate layers. \Tian{This variance in contribution across layers not only affects performance but also system efficiency, facilitating tailored adaptations for edge deployment. To meet heterogeneous hardware constraints, \model~ redefines LLM pruning as a generative task, surpassing traditional manually designed discrete heuristic optimizations to identify and remove non-essential structures in a system-efficient manner. }Specifically, it employs a data-driven tailor that identifies the optimal pruning configuration through gradient-based optimization within a continuous representation space. As illustrated in Figure~\ref{fig:offline}, the generative tailor comprises four key modules: 

1) \textit{``Ratio-score'' Data Collection.} 
Considering the memory constraints of the edge device and the dimensions of the model to be deployed, we initially establish the overall reduction ratio for the LLMs. For fine-grained optimization, we develop an exploration-exploitation strategy to determine layer-wise level pruning ratios $r_{i}$. This approach utilizes heuristic-based pruning methods~\cite{LLM-Pruner,ShortGPT,SliceGPT} to produce high-quality ratios and incorporates random pruning ratios as part of the exploration process.
In contrast to prior works \cite{hubara2018quantized,importance_pruning,Channel_pruning} primarily focus on reducing the model size and computational overhead, \model~ integrates broader criteria including generation ability, inference latency, and energy cost. We define a holistic metric function $f$ to characterize overall performance for a given $r_{i}$, aiming for optimization suitable for resource-constrained edge devices:
\begin{equation}
    \scalemath{0.95}{
    \textcolor{black}{s_{i} =  \textit{f} (r_{i})=\frac{1}{ppl_{i}} \times\left(\frac{E}{e_i}\right)^{\textbf{1}\left(E<e_i\right) \times \alpha}
    \times\left(\frac{T}{t_i}\right)^{\textbf{1}\left(T<t_i\right) \times \beta}}},
    \label{equation:metric}
\end{equation}
where $ppl$ is the zero-shot perplexity~\cite{ppl-1}, quantifies the generative capabilities, with lower values indicating more precise model predictions. $T$ and $E$ denote the latency and energy budgets specific to edge, respectively. $t_{i}$ and $e_{i}$ represent the latency and energy consumption for a given ratio $r_i$. $\textbf{1}(x)$ is an indicator that returns 1 if condition $x$ holds, and 0 otherwise. Thus, configurations that exceed the thresholds $T$ and $E$ are penalized by developer-specified factors $\alpha$ and $\beta$, both set to 2 in our implementation. Finally, we obtain comprehensive ratio-score pairs, denoted as $\mathcal{P} = {(r_{i}, s_{i})}$.

2) \textit{Continuous Space for Tailoring.} Traditional approaches \cite{LLM-Pruner,ShortGPT,SliceGPT} often employ discrete pruning spaces, resulting in heuristic and incomprehensive pruning ratios. In order to characterize the continuous tailoring optimization space of edge devices, we implement an encoder-evaluator-decoder framework. This architecture includes a single-layer LSTM network \cite{hochreiter1997long} functioning as both encoder and decoder, alongside a feed-forward neural network serving as the evaluator. This setup effectively embeds ratio-score pairs $(r_{i}, s_{i})$ into a continuous representation space $\Theta$.

3) \textit{Gradient-based Tailoring Optimization.}
With the well-trained representation space $\Theta$, we utilize a gradient-based optimization method to identify the optimal pruning configuration. We select top-$K$ collected pairs serving as starting points to ensure effective initialization. Denoting such starting points as $\mathcal{E}{r}$, we initiate the optimization process along the gradient direction driven by the evaluator $\pi$:
\begin{equation}
\begin{aligned}
\scalemath{0.95}{
\mathcal{E}_{r}^{*}= \mathcal{E}_{r} + \eta \frac{\partial \pi (\mathcal{E}_r)}{\partial \mathcal{E}_r}},
\end{aligned}
\label{equation:gradient}
\end{equation}
where $\mathcal{E}_{r}^{*}$ denotes the optimal pruning configuration representation, and $\eta$ is step size used in the gradient update.

4) \textit{Optimal Tailoring Generation.}
Based on the optimal pruning representation $\mathcal{E}_{r}^{*}$, the optimal pruning configuration $r^{*}$ is identified by the trained decoder $\xi$, denoted as $r^{*} = \xi(\mathcal{E}_{r}^{*})$. To generate $r^{*}$ iteratively without pre-specifying the ratios, we employ a beam search strategy~\cite{freitag-al-onaizan-2017-beam}, allowing for a systematic exploration of potential configurations to achieve the best possible performance. The generation process continues until the stop token $\langle$EOS$\rangle$ is encountered, enhancing the adaptiveness of the model configuration. Finally, remove non-essential groups through structural pruning, guided by the generated optimal pruning ratio.

According to \textit{\newnew{Motivation} 2}, the patterns of user application usage and the inter-dependencies among tasks offer opportunities for optimization in downstream tasks, which can reduce the costs of fine-tuning and improve model performance. Therefore, plug-and-play LoRA adapters are utilized to enhance the tailored LLM generalization capabilities across diverse downstream applications.
\Tian{For \( n \) downstream applications, a set of LoRAs, \(\Phi = \{\phi_1, \phi_2, \ldots, \phi_n\}\), is initialized, where each LoRA \(\phi_i = BA\) is trained per task. The forward computation is:\(y' = y + \Delta y = \mathbf{W}_{0}x + BAx,\) where \( y' \in \mathbb{R}^d \) is the output, \( x \in \mathbb{R}^k \) is the input, \( B \in \mathbb{R}^{d \times r} \), \( A \in \mathbb{R}^{r \times k} \) with \( r \ll \min(d, k) \), and \( \mathbf{W}_{0} \) are frozen LLM parameters. \( B \) is initialized to zero and \( A \) with Gaussian values.}

\begin{figure*}[!t]
    \centering
    \subfigure[Software]{
\includegraphics[width=0.48\linewidth]{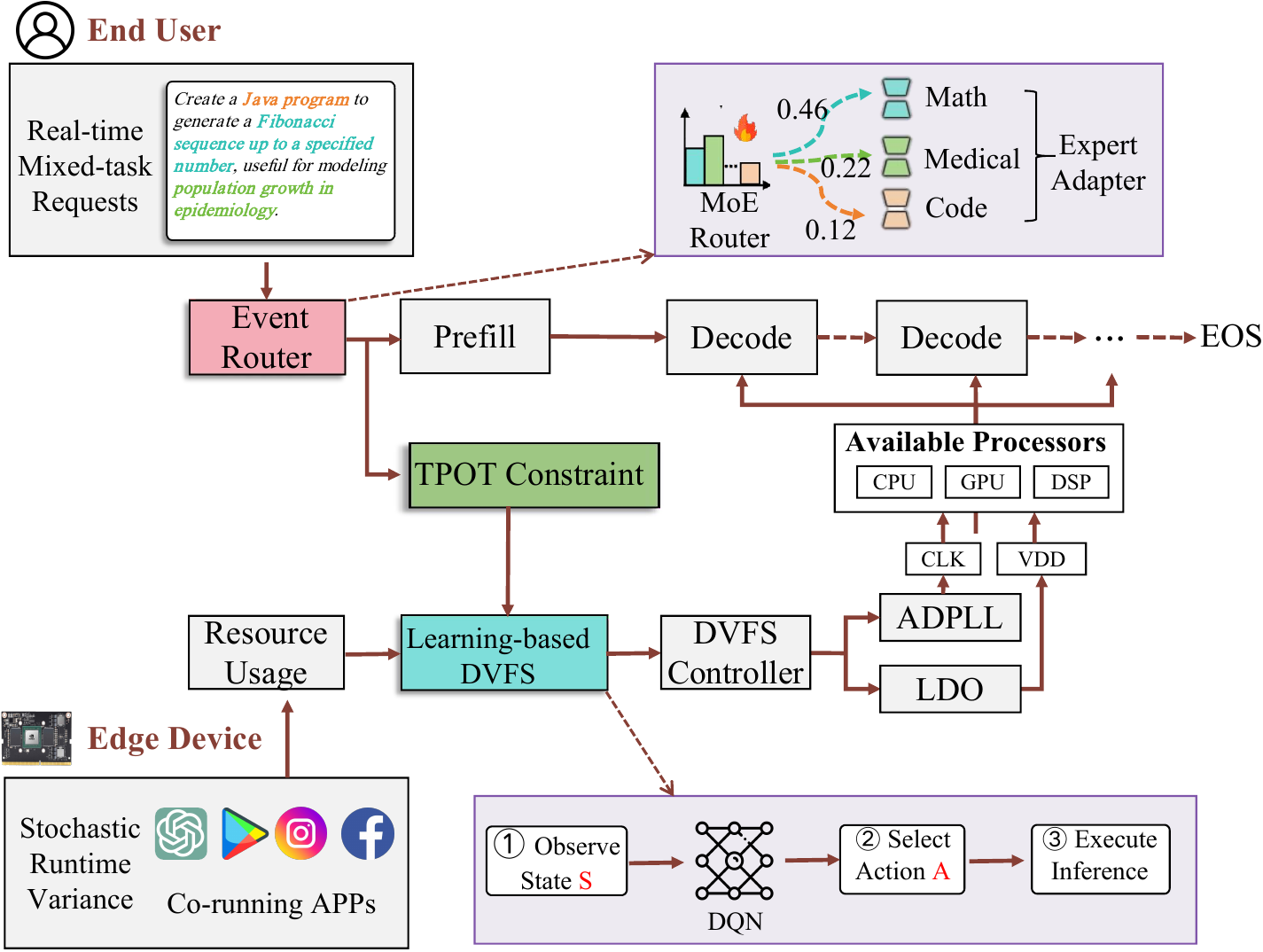}}
    \subfigure[Hardware]{\includegraphics[width=0.48\linewidth]{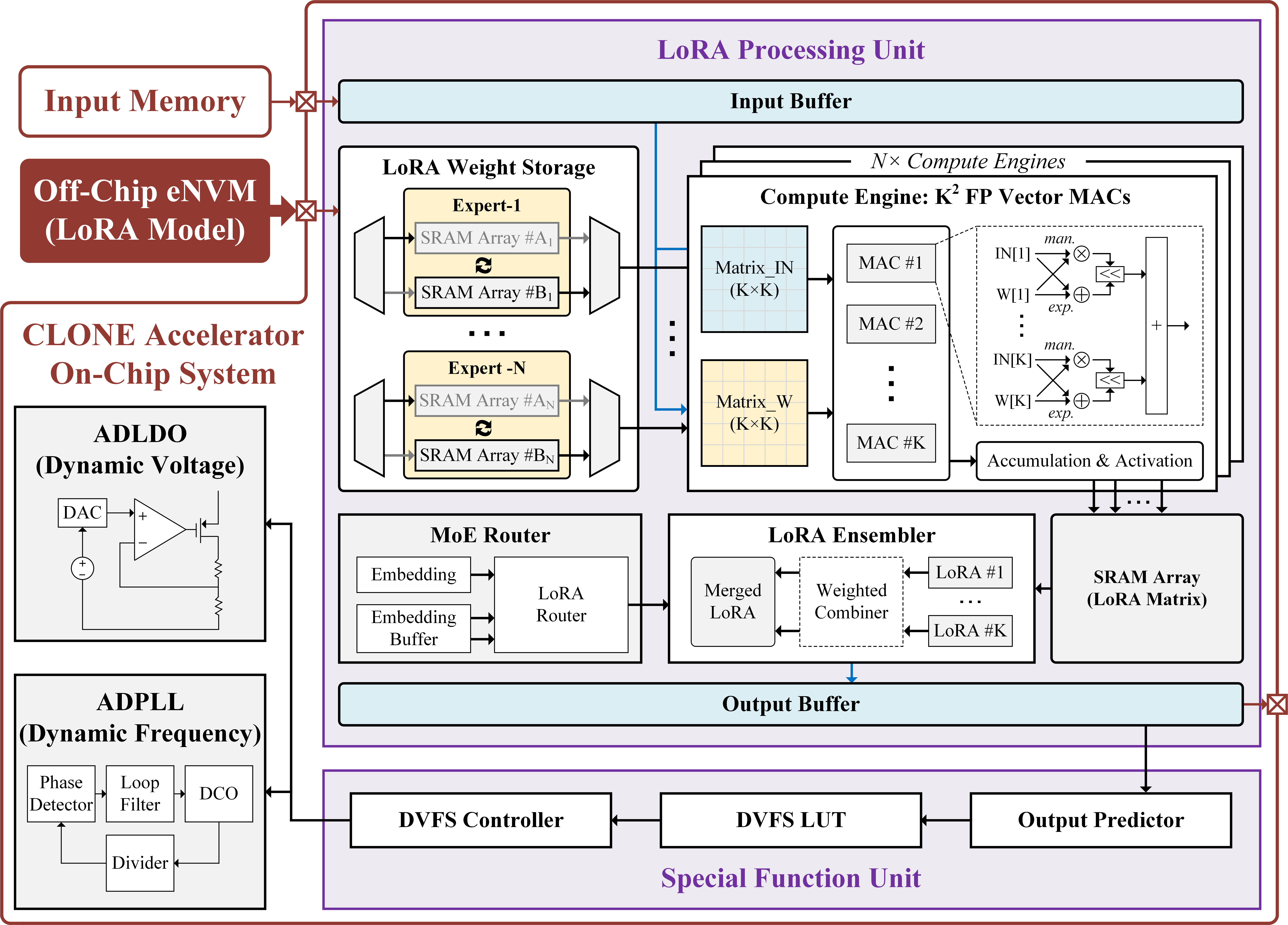}}
    \caption{Online latency-aware inference workflow. \textbf{Left (Software)}: \model~ employs a request-wise MoE router for dynamic LoRA configuration to manage layer-wise DVFS, optimizing energy efficiency within the specified latency targets. \textbf{Right (Hardware)}: The hardware accelerator system features a request-wise LoRA Processing Unit (LPU) designed for LoRA adaptation, and a Special Function Unit (SFU) tailored for pre-token DVFS at token layer boundaries.}
    \label{fig:online-dvfs}
\end{figure*}

\subsection{Online Latency-aware Inference}
\textit{\newnew{Motivation} 2 \& 3} reveals that current approaches to real-time inference are constrained to workload-level resource adjustments, resulting in energy wastage. It identifies a crucial potential for system optimization via fine-grained management of performance and energy.
\Tian{
While offline device-specific tailoring effectively reduces memory constraints and enhances downstream performance, the achieved latency can fluctuate significantly due to diverse user requests and runtime variability, potentially violating real-time latency constraints and increasing energy consumption. As shown in Figure~\ref{fig:online-dvfs}, at the model level, a MoE-based router dynamically selects the optimal LoRA adapter based on stochastic user prompts, improving the accuracy of LLM-powered application responses. At the system level, \model~ integrates a token predictor to estimate the output token count, guiding a learning-based DVFS controller to adjust frequency and voltage for each token at layer boundaries. This minimizes per-request energy consumption while adhering to real-time latency targets. Additionally, \model~ embeds these efficient algorithmic enhancements into a 28nm on-chip accelerator architecture.}

\noindent \textbf{Request-wise MoE-based Router.}
In practice, end-user requests cover a diverse range of prompts, each associated with different tasks. Consequently, as shown in Figure \ref{fig:online-dvfs} (a), an MoE (Mixture-of-Experts) \cite{moe,ShazeerMMDLHD17} router has been developed to dynamically and effectively merge LoRA modules for each mixed-task prompt. This development significantly extends the plug-and-play capabilities of \model~. A set of experts $\textbf{E} = (E_{1}, E_2, \ldots, E_{N})$, where each expert $E_{i}$ represents a tuned LoRA module. For input $x_{i}$, the output $y_i$:
\begin{equation}
    y_i 
     =\mathbf{W_{o}} x_i+\sum_{j=1}^N \omega_{ij} \cdot E_j\left(x_i\right),
\end{equation}
where $\omega_{ij}$ modulates the contribution weights of each expert. Unlike traditional methods~\cite{LoRAMoE_1, tian2024hydralora} that incur extra computation and storage from trainable gate functions, we propose a parameter-free soft MoE method leveraging dynamic prompts. Using a sentence-embedding model, $\Gamma$ (here adopting BGE \cite{bge-m3}), the input embedding is formulated as $\Gamma(x)$. For each LoRA module $\phi$, the embedding is obtained from randomly selected domain-specific samples $k$, expressed as $\Gamma(\phi) = \frac{1}{k} \sum_{i=1}^{k} \Gamma(k_{i \cdot \phi})$. To measure the similarity between the LoRA module $\phi$ and the prompt $x$, we leverage the cosine similarity, denoted as $\sigma$, as follows:
\begin{equation}
    \sigma(x,\phi) = cos(\Gamma(x),\Gamma(\phi)),
\end{equation}
Followed by a softmax function which takes an intermediate token representation as input and combines the output of each expert based on the gating weight
$\Omega = (\omega_1, \dots, \omega_N)$:
\begin{equation}
    \Omega = \text{softmax}(\textbf{s}_x),
\end{equation}

\noindent \textbf{Learning-based DVFS Controller.} 
\Tian{To further improve system effectiveness for LLM inference at the edge, a learning-based DVFS is developed to reduce per-generated token energy consumption while satisfying the real-time latency target at the layer-wise level. Among various reinforcement learning (RL) methods~\cite{RL_1,RL_4,Q_table,RL_2,RL_3}, \model~ employs a simple two-layer MLP to efficiently learn policies within an episodic Markov decision process, ensuring minimal latency overhead. RL involves three core components:} 

$\textbf{\textit{State}}$: \Tian{Edge inference efficiency is heavily influenced by the processor intensity of co-running applications, denoted as \( S_{pro} \). For runtime SLO constraints, inference efficiency is tightly coupled with frontend prefill time TTFT (\( T_{PRE} \)) and per-token decoding latency TPOT target (\( T_{DEC} \)).}

$\textbf{\textit{Action}}$: In RL, actions represent the adjustable control parameters of the system. For LLM inference at the edge, we define these actions as the selectable execution settings, specifically the energy-optimal supply voltage ($V_{DD}$) and the optimal running frequency ($F_{req}$) for each layer. Provided that the QoS constraints are met, it is feasible to lower the processor frequency, thereby conserving energy.

$\textbf{\textit{Reward}}$: In RL, a reward models the optimization objective of the system. We encode energy optimization within the execution target as the reward, $\mathbb{R}_{energy}$, which is calculated using the power model \cite{power_1,power_2}:
\begin{equation}
    \scalemath{0.9}{\textcolor{black}{\mathbb{R}_{energy} = \sum_{f}(P^{f}_{DEC} \times T_{DEC}  + P^{f}_{PRE} \times T_{PRE})}}
\end{equation}
where $P^{f}_{PRE}$ and $P^{f}_{DEC}$ , representing the power consumption of the CPU/GPU at each frequency during the prefill and decoding states respectively, are determined through power measurements. These values are subsequently stored in a lookup table (LUT) within \model~ for efficient retrieval.

\subsection{Hardware Accelerator System.} 
\new{As illustrated in Figure~\ref{fig:online-dvfs} (b), acknowledging the limitations inherent in purely software-based approaches, \model~ introduces a scalable 28nm hardware accelerator specifically designed for edge-based LLM inference, effectively translating software optimizations into concrete performance improvements.} This system integrates a LoRA Processing Unit (LPU) for dynamic adapter hot-swapping, enabling adaptive model performance optimization via dedicated data paths. Complementing this, a Special Function Unit (SFU) is equipped to perform continuous, fine-grained DVFS adjustments. The LPU executes a request-wise MoE router algorithm to efficiently manage stochastic requests, mitigating the limitations of general-purpose processors and enhancing computational throughput. Unlike conventional LoRA weight stored in on-chip SRAM, which either require frequent reloads from DRAM during wake-up cycles or keep SRAM active, wasting leakage power~\cite{eNVM,eNVM_2,eNVM_3}, we utilize an embedded non-volatile memory (eNVM) buffer to retain LoRA modules, eliminating reload overhead upon power-on. Simultaneously, the SFU employs a learning-based DVFS mechanism to adapt to runtime variability, ensuring rapid convergence to the desired voltage (\( V_{DD} \)) and frequency (\( F_{req} \)) settings. The SFU employs a lightweight predictor and DVFS model implemented as lookup tables (LUTs) for efficient hardware operation and action determination. The generated voltage (\(V_{DD}\)) and frequency (\(F_{req}\)) are realized using a fast-switching low-dropout (LDO) voltage regulator~\cite{LDO,LDO_2} and an all-digital phase-locked loop (ADPLL)~\cite{ADPLL,ADPLL_2}, ensuring rapid runtime adjustments. Communication between the LPU and SFU is facilitated via a custom-built bi-directional streaming channel. An AXI splitter arbitrates the control flow of instructions and data bound for the LPU and SFU AXI-slave partitions.

\section{EVALUATION}

\subsection{Experimental Setup}
\noindent \textbf{Infrastructure.} In order to comprehensively evaluate the efficiency and effectiveness of \model~, We perform our experiments on two heterogeneous edge devices: Orin Nano~\cite{nano} and Jetson Orin NX \cite{nvidia_jetson_orin}. Table \ref{tab:device} summarizes their specifications. To emulate heterogeneous deployment models, \textcolor{black}{we utilize Llama-7B \cite{llama}, Llama2-7B \cite{Llama_2}, Llama2-13B and Vicuna-7B \cite{vicuna}} as baseline models. To simulate the heterogeneous performance requirements of multiple tasks at the edge, we employ the Flanv2 \cite{flanv2} dataset, which comprises 46 tasks across 10 domains, serving as the edge downstream dataset. Devices run stochastic web-search to simulate runtime variance. CodeCarbon\cite{codecarbon} is utilized to measure runtime duration and the power consumption.

\noindent \textbf{Baselines.}
To evaluate the effectiveness of \model~, we compare \model~ with 7 representative approaches, including: 
\textbf{\textit{A. Vanilla:}}
(1) \textit{Vanilla} assess the effectiveness of the original LLMs as the baseline criterion.
\textbf{\textit{B. Model compression:}}
(2) \textit{Random} randomly selects certain groups for pruning.
(3) \textit{LLMPruner} \cite{LLM-Pruner} removes connected structures of dependent multihead attention from LLMs based on gradient information.
(4) \textit{ShortGPT} \cite{ShortGPT} defines the Block Influence metric to guide redundant subset removal.
(5) \textit{SliceGPT} \cite{SliceGPT} replaces each weight matrix in the network with a smaller, dense matrix, thereby reducing the network embedding dimension. 
\textbf{\textit{C. Computation optimization:}}
(6) \textit{FlexGen} \cite{flexgen} retains all model parameters on CPU DRAM. During inference, the demand weights are loaded from the CPU to the GPU.
\textit{\textbf{D. Small size model:}}
(7) OpenLLaMA-3B\cite{openlm2023openllama}, an open reproduction of Llama with 3B parameters, trained on 1T tokens.

\begin{figure}[!t]
    \centering
    \includegraphics[width=0.85\linewidth]{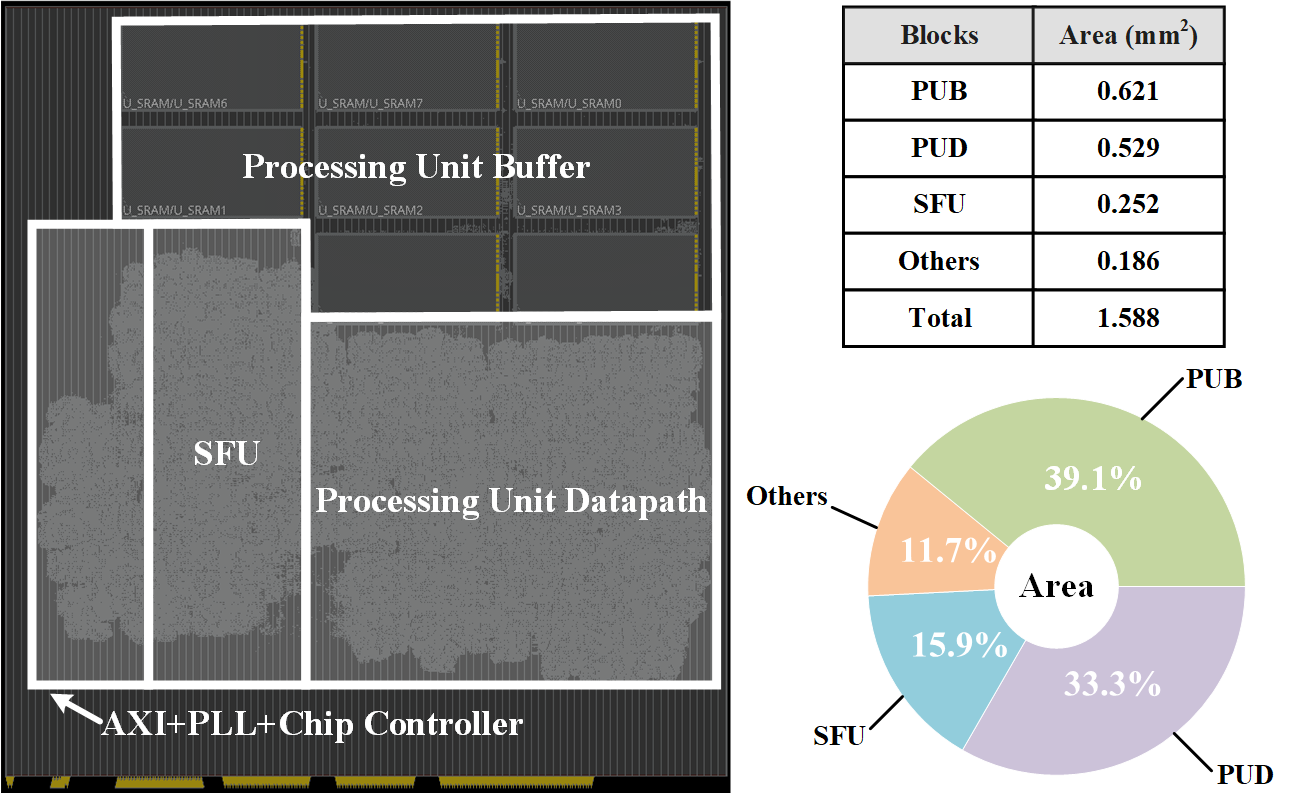}
    \caption{28nm physical layout and area breakdown of the energy-optimal \model~ accelerator system.}
    \label{fig:layerout}
\end{figure}

\begin{figure}[!t]
    \centering
    \includegraphics[width=0.85\linewidth]{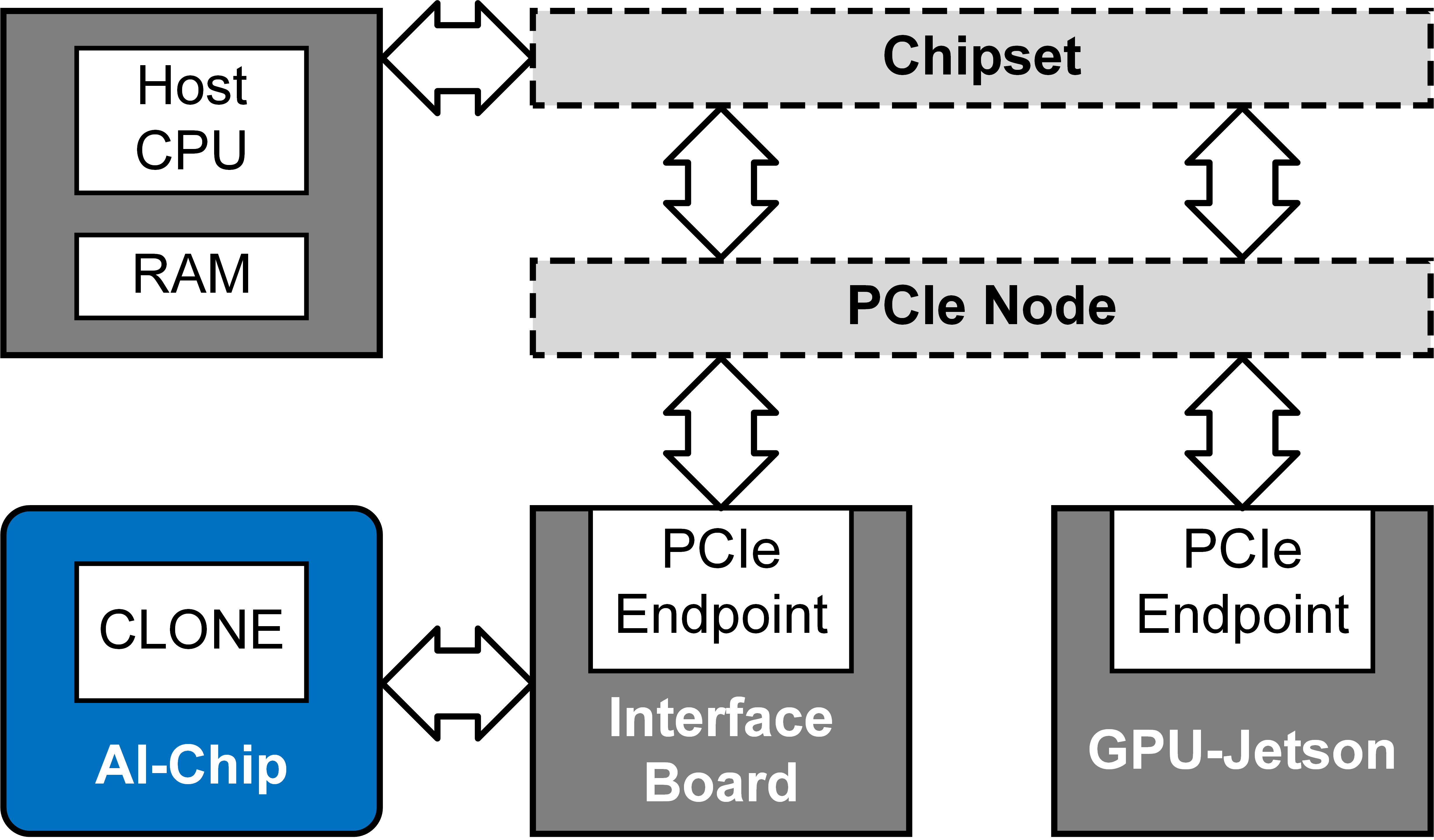}
    \caption{\textcolor{black}{Test system. Integration of the \model~ accelerator into the Jetson platform via a PCIe interface to facilitate high-performance data exchange.}}
    \label{fig:\model~_dataflow}
\end{figure}

\noindent \textbf{Evaluation Metrics.}
In order to comprehensively evaluate the efficiency and effectiveness of \model~, we conduct evaluations from three perspectives. \textit{\textbf{A.Generation ability.}} We measure the zero-shot perplexity (PPL) analysis on WikiText2 \cite{Wikitext2} and PTB \cite{Ptb}, two of the most commonly used evaluation datasets, to evaluate the generation capabilities of the customized model. Lower PPL values signify stronger generation ability.
\textit{\textbf{B. General-purpose task solving ability.}} To demonstrate the world knowledge and problem-solving skills of the customized model, we utilize three benchmarks: 
1) BBH \cite{bbh}, which includes 23 challenging tasks such as Q$\&$A, natural language reasoning, and sentiment analysis.
 2) MMLU \cite{MMLU}, covering 57 tasks across diverse domains like mathematics, history, law, and ethics.
3) Commonsense reasoning tasks used in Llama paper \cite{llama}, featuring BoolQ \cite{boolq}, PIQA \cite{piqa}, OBQA \cite{OpenbookQA}, ARC-c \cite{arc}, ARC-e \cite{arc}, WinoGrande \cite{WinoGrande}, and HellaSwag \cite{hellaswag}.
\textit{\textbf{C. System effectiveness.}} We evaluate the system effectiveness of \model~ from two perspectives: latency and energy consumption, which are defined as the duration and energy cost to infer the WikiText2\cite{Wikitext2}.

\noindent \textbf{Hyperparameter and Reproducibility.}
\Tian{For offline device-specific tailoring, we first execute classic approaches for 100 epochs to collect training data, and use $25\times$ randomly shuffled client selection as data augmentation for tailoring training. Both Encoder and Decoder use a single-layer LSTM configuration, while the Predictor has a dual-layer feed-forward setup. Hidden state dimensions are 64 for Encoder and Decoder, and 200 for the Predictor, with all layers having an embedding size of 32. The hyperparameters include a batch size of 1024, a learning rate of 0.001, and $\eta=0.8$. For optimization, we start with the top 25 model pruning ratio records. The LoRA rank $r$ and the scaling hyperparameter $\alpha$ are set to 8 and 16, respectively, with 3 training rounds.}


\subsection{\textcolor{black}{Test System}}
\new{Figure \ref{fig:layerout} presents the layout of the energy-optimal 28nm \model~ accelerator system. After place‑and‑route, the core footprint is only 1.588 mm$^{2}$; the figure also delineates module boundaries and reports the post‑layout power‑ and area‑breakdowns. 
\newnew{Due to the high cost and time requirements of a full tape-out}, the \model~ accelerator was validated through post-layout simulation. The physical layout was generated from Cadence Innovus synthesis results, following a complete design flow: behavioral modeling, RTL design, synthesis, P$\&$R, and physical verification, ensuring simulation fidelity to actual fabrication.}
Figure \ref{fig:\model~_dataflow} illustrates the integration of the \model~ accelerator into the Jetson platform via a PCIe interface to facilitate high-performance data exchange. The data flow initiates with the AI-Chip (\model~), which interfaces with the Jetson GPU through a PCIe connection. \new{An interface board bridges the PCIe endpoints of the CLONE accelerator and the Jetson GPU}. This architecture is supported by the host CPU and RAM, where the CPU handles top-level data flow control, ensuring that requests are efficiently managed and routed. The \model~ is responsible for executing tasks based on user inputs, such as selecting the most efficient LoRA group and performing edge real-time DVFS control to optimize performance. By bypassing system software dependencies, this hardware-centric setup ensures seamless communication and data transfer between the \model~ and the Jetson device, achieving high levels of performance and energy efficiency.

\begin{figure}[!t]
    \centering
    \includegraphics[width=0.95\linewidth]{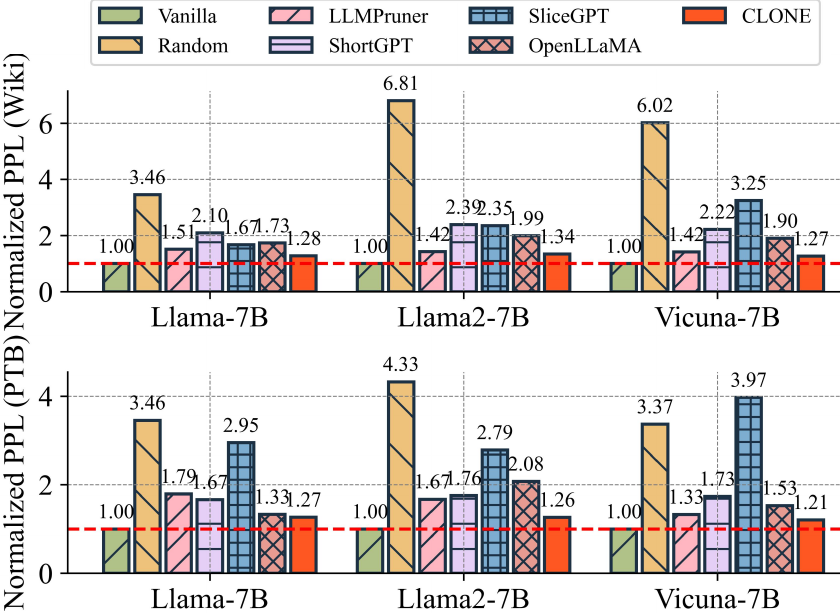}
    \caption{Generation ability comparison of various schemes on WikiText2 and PTB dataset across 3 representative LLMs.}
    \label{fig:eval_PPL}
\end{figure}

\begin{figure*}[!ht]
    \centering
    \includegraphics[width=\linewidth]{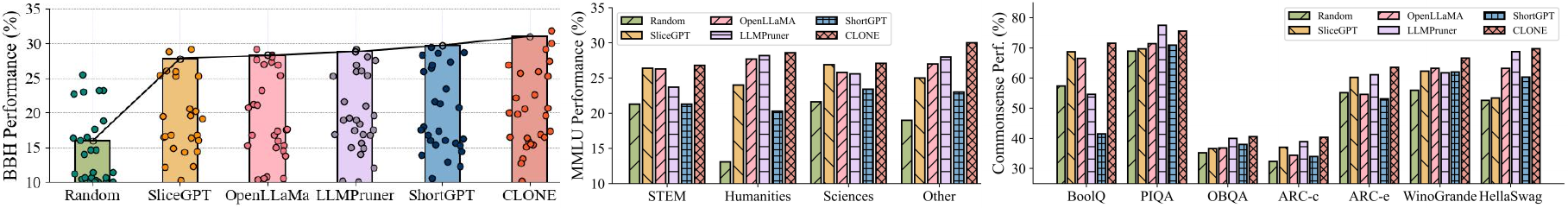}
    \caption{Downstream task performance of different customization approaches of Llama-7B on 3 representative benchmarks, including BBH (zero-shot), MMLU (3-shot), and Commonsense (zero-shot).}
    \label{fig:eval_perf}
\end{figure*}

\subsection{Evaluation Results and Analysis}
\noindent \textbf{Generation Ability.} 
To evaluate the customized model generation ability, we use the WikiText2 and PTB datasets to measure the zero-shot PPL (lower values correspond to stronger text generation capability) across different models on the Orin NX. Figure~\ref{fig:eval_PPL} shows the zero-shot PPL values that compare \model~ with the baselines. In order to effectively demonstrate the characteristics of different schemes, we normalize the measured PPL values to the vanilla model. Compared to baselines, \model~ preserves the most generation ability, achieving up to $5.1\times$ generating capacity of \textit{Random} on WikiText2 and up to $3.4\times$ on PTB. 
This can be attributed to \model~ using PPL as a pruning metric to construct a continuous pruning selection space. It then utilizes gradient-based optimization to identify the optimal pruning configuration that minimally impacts generation ability.

\begin{figure*}
    \centering
    \subfigure[Llama2-7B]{\includegraphics[width=0.33\linewidth]{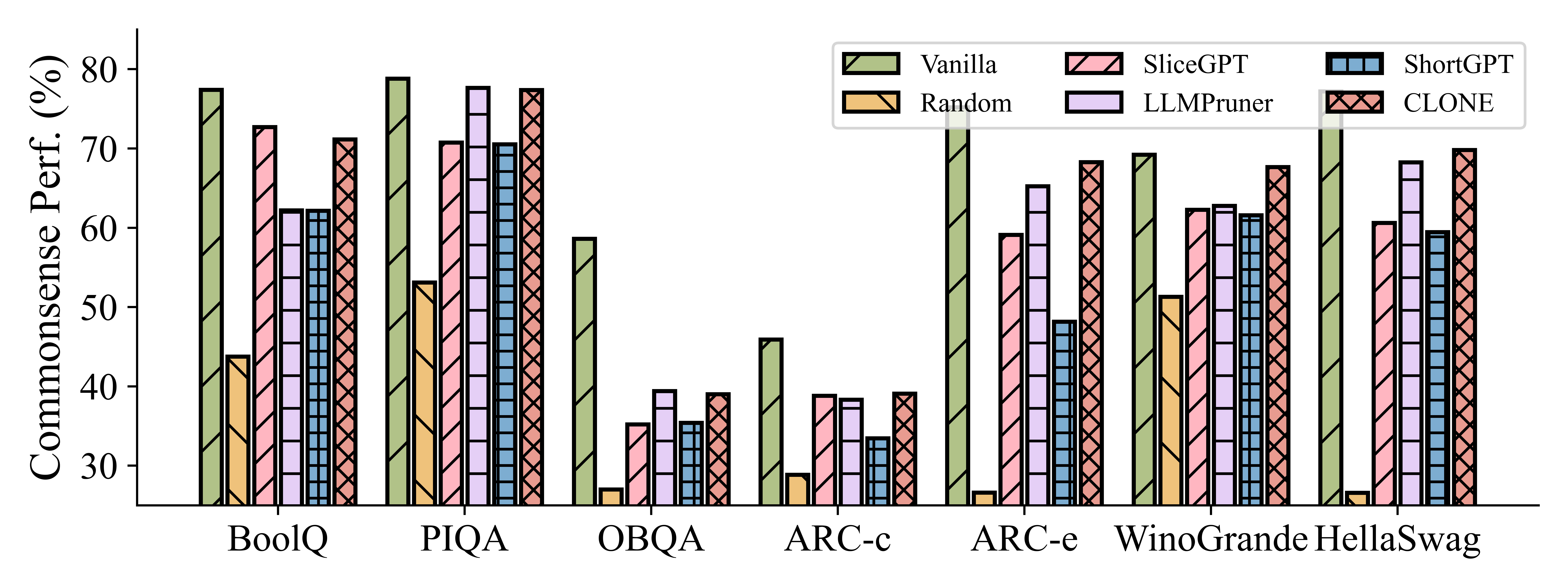}}
    \subfigure[Vicuna-7B]{\includegraphics[width=0.33\linewidth]{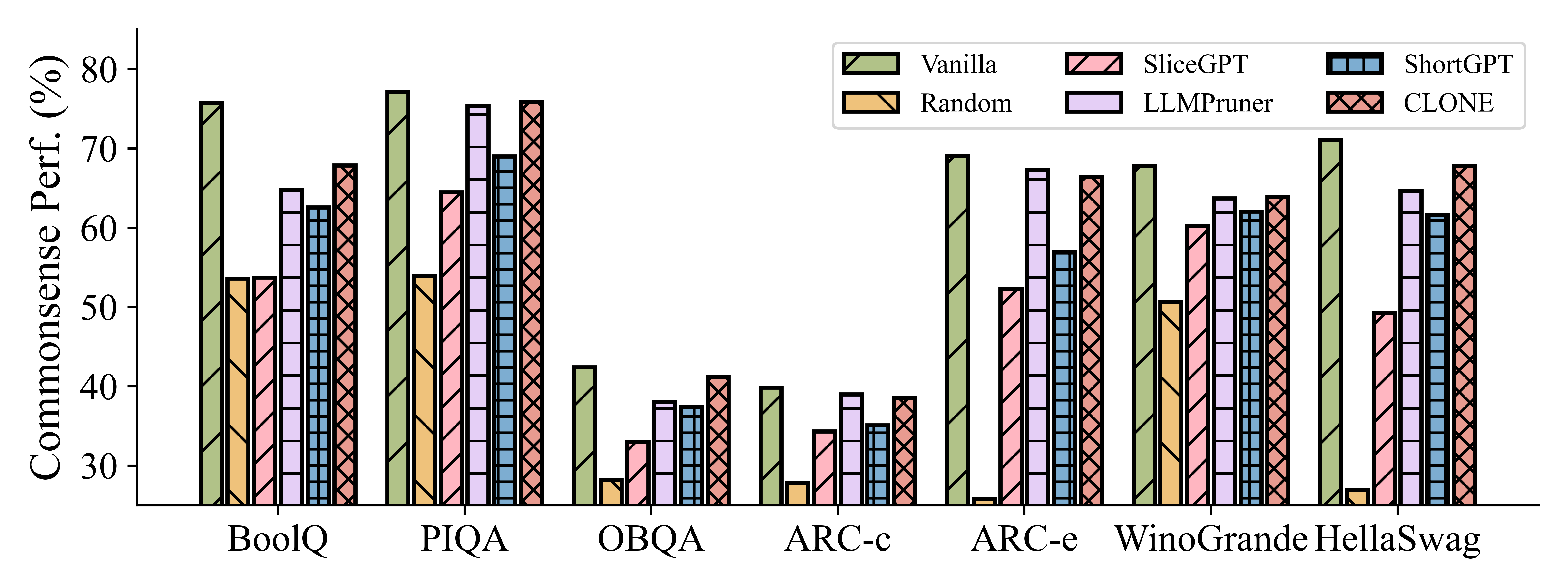}}
    \subfigure[Llama2-13B]{\includegraphics[width=0.33\linewidth]{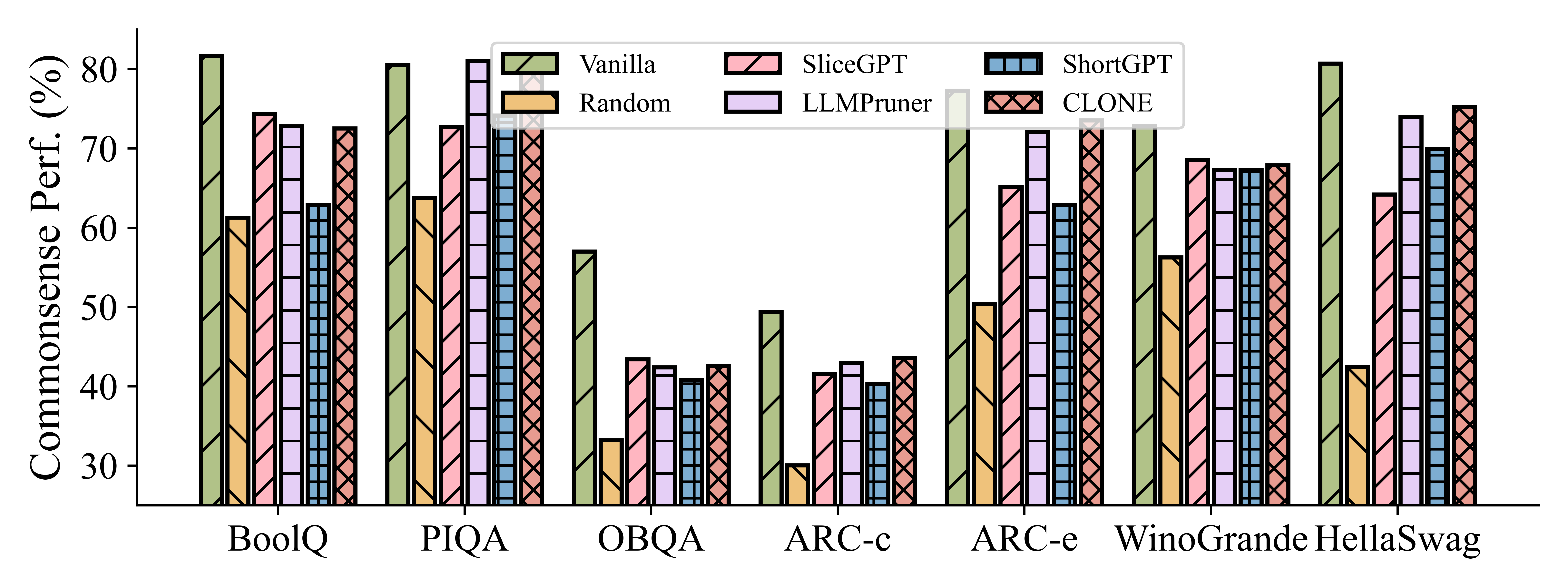}}    
    \caption{\new{Zero-shot performance (commonsense) of Llama2-7B, Vicuna-7B and Llama2-13B.}}
    \label{fig:diff_model}
\end{figure*}

\noindent \textbf{Downstream Task Performance.}
\textcolor{black}{Figure~\ref{fig:eval_perf} illustrates the downstream task performance of \model~ across three representative benchmarks (BBH, MMLU, and Commonsense), covering 87 tasks. We observe that \model~ significantly outperforms other baselines across various domains and tasks. Specifically, for BBH, \model~ improves the average accuracy by 15.1\% over Random and by 2.37\% over others on average; for MMLU, it increases average accuracy by 6.0\% over Random and by 2.96\% over others on average; for commonsense tasks, \model~ enhances average accuracy by 10.1\% over Random and by 6.1\% over others on average. The principle as discussed in $\S \ref{section:opportunity}$, \model~ implements task-aware tailoring for specific devices by considering end-user usage patterns and task specificity and relevance. It selects appropriate types and quantities of LoRA for fine-tuning. Moreover, we note that baseline performance varies across different downstream tasks. For instance, in BBH, ShortGPT is the sub-optimal model, whereas in Commonsense, LLMPruner is the sub-optimal. This variability stems from the heuristic and incomplete nature of baseline approaches. In contrast, \model~ utilizes a request-wise MoE-based router to effectively manage the properties of diverse real-world requests, resulting in superior LoRA module fusion. Overall, this experiment demonstrates \model~'s practical and robust ability to scale to complex workloads and applications.}

\noindent \textbf{Robustness and Scalability.}
\textcolor{black}{Figure~\ref{fig:diff_model} details the zero-shot performance of the pruned model across a variety of downstream tasks, as well as its zero-shot perplexity on the WikiText2 and PTB datasets, utilizing diverse pruning configurations. Our observations confirm that \model~ consistently outperforms established baselines across all metrics.\textit{1) Different types of models}: Comparing Llama2-7B and Vicuna-7B, the results indicate that \model~ exhibits a strong generalization capability, outperforming baselines by an average of 23.85\%. \textit{2) Different model sizes}: Evaluating Llama2-7B and Llama2-13B, it is evident from Figure~\ref{fig:diff_model} that \model~ maintains 91.13\% of the performance of the unpruned (vanilla) models, underscoring its scalability. These results collectively affirm that \model~ is scalable to different LLMs in general, not limited to specific LLMs, so it is not a one-off ``software-ASIC''.}

\begin{figure}[!ht]
    \centering
    \includegraphics[width=0.9\linewidth]{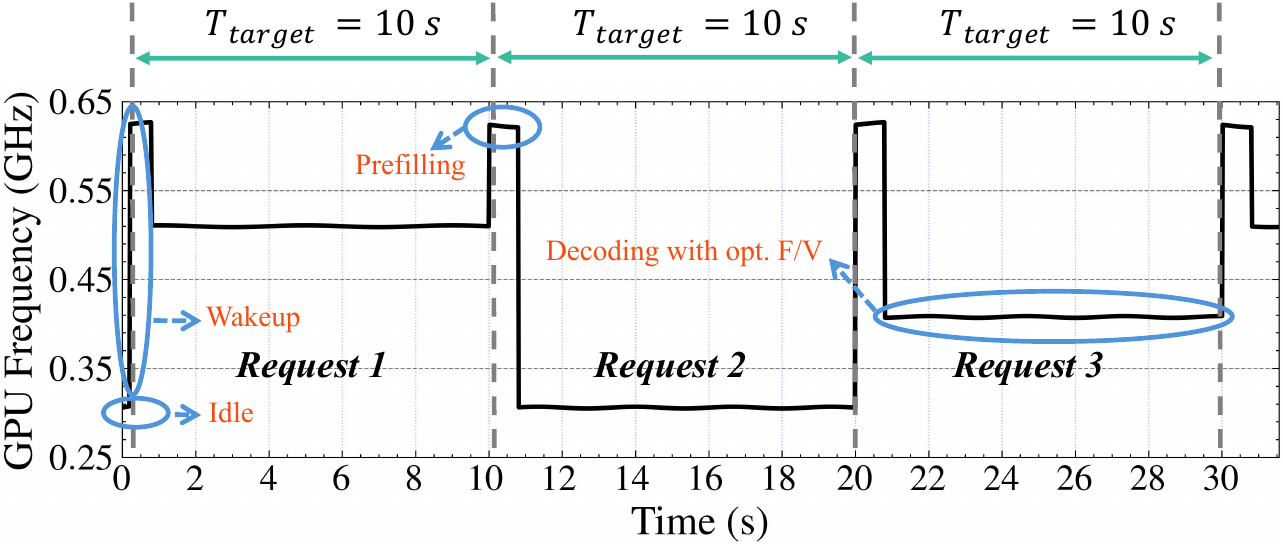}
    \caption{Simulations of LDO dynamic $V_{DD}$ adjustments.}
    \label{fig:LDO}
\end{figure}




\noindent \textbf{System Effectiveness.}
Then, we evaluate the system efficiency of \model~ from latency and energy consumption during the LLMs inference process on the WikiText2 dataset at the off-the-shelf devices Orin NX and Nano. As shown in Table~\ref{tab:sys_result}, we can see that \model~ effectively speeds up the inference process up to $11.92\times$, and achieves significant energy saving up to $7.36\times$.
Figure~\ref{fig:LDO} shows the spice-level simulation of the DVFS for runtime inference.
There are two possible reasons 1) As shown in Equation~\ref{equation:metric}, \model~ jointly considers the latency and energy budgets as metrics to generate optimal pruning configuration. 2) \model~ decouples prefill and decoding phases, applying request-level DVFS to boost effectiveness. In contrast, FlexGen trades high latency and energy for full-parameter LLMs, while methods like LLMPruner and ShortGPT focus on static, one-off compression without considering system-level optimization.



\begin{table}[!ht]
    \centering
 \caption{\new{System efficiency (latency and energy) comparison of various schemes to inference WikiText2 dataset.}}
    \label{tab:sys_result}
    \resizebox{0.95\linewidth}{!}{\begin{tabular}{c|cc|cc}
    \bottomrule[1.5pt]
        \rowcolor{mygray} & \multicolumn{2}{c|}{\textbf{Energy (Wh)}} & \multicolumn{2}{c}{\textbf{Latency (s)}} \\
        \rowcolor{mygray} \multirow{-2}{*}{\textbf{Method}} & \textbf{Jetson NX} & \textbf{Jetson Nano} & \textbf{Jetson NX } & \textbf{Jetson Nano} \\
         \toprule[0.75pt]
       Random & 7.27 & 8.26& 842.40 & 1145.07\\
       SliceGPT & 5.47 & 7.54& 661.65&929.39 \\
       OpenLLaMA & 5.67 &7.70 & 506.73& 662.73\\
       LLMPruner & 6.01 &6.91 & 622.92&1023.51 \\
       ShortGPT & 5.67 &8.56 &555.14 &698.02 \\
       FlexGen & 21.12 &26.04& 3166.27 &4674.42 \\ \hline
       \new{\model~$^{-HW}$} &4.81 & 5.56 & 462.72 & 552.18 \\
       \textbf{\model~} &\textbf{3.46} & \textbf{3.54}&\textbf{322.76} & \textbf{392.15} \\
    \bottomrule[1.5pt] 
    \end{tabular}}
\end{table}

\begin{figure}[!t]
    \centering
    \includegraphics[width =0.85\linewidth]{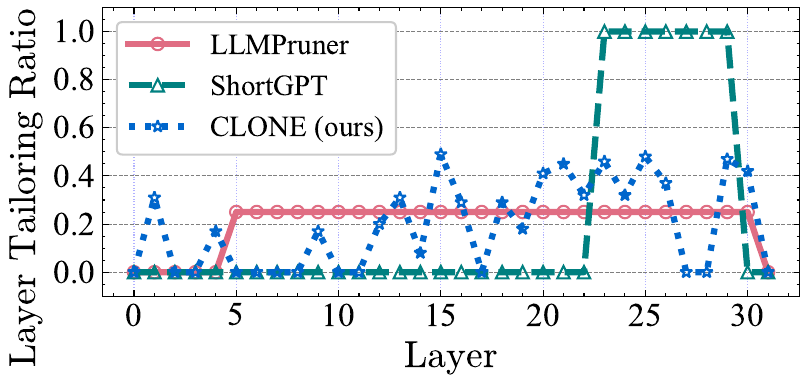}
    \caption{Comparison of the model pruning configuration.}
    \label{fig:ratio}
\end{figure}

\noindent \textbf{Study of the Generative Pruning Configuration.}
To examine how \model~ adaptively generates the optimal model pruning configuration, we compare the generative 
 pruning approach to the layer-wise method used by ShortGPT and the uniform direct average method employed by LLMPruner on Llama-7B to illustrate the strategic disparities in their decision-making processes. As presented in Figure~\ref{fig:ratio}, LLMPruner remains static from the 5th to 30th layer. In contrast, ShortGPT uses only binary values of 0 or 1. We note that the layer-level pruning configuration determined by \model~ dynamically varies across layers. This adaptability reflects a more nuanced decision-making process, where the contributions of individual layers are recognized as heterogeneous, as discussed in \textit{\newnew{Motivation} 1}. This indicates that \model~'s generative decision-making is more adaptive and tailored to the specific contributions of each layer. 

\noindent \textbf{\new{Effectiveness of the Hardware Accelerator.}}
\new{The accelerator boosts efficiency by exploiting dedicated hardware resources; it does not itself alter generation quality or task‑specific accuracy. Disabling it forces the workload onto general‑purpose compute units, eliminating these hardware‑level gains. As Table~\ref{tab:sys_result} shows, energy rises from 3.46Wh to 4.81Wh and end‑to‑end latency from 322.76s to 462.72s without the accelerator \model~$^{-HW}$. Even in this regime, our offline, metric‑guided generative pruning (Eq.~\ref{equation:metric}) still surpasses all baselines, underscoring its robustness.}

\begin{figure}[!t]
    \centering
    \subfigure[w/o MoE]{\includegraphics[width =0.315\linewidth]{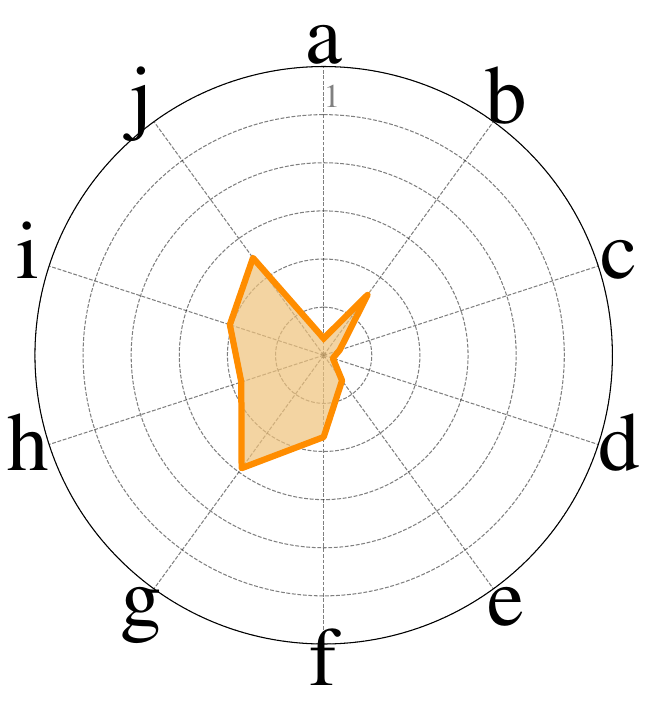}}
    \subfigure[MoE (Top-1)]{\includegraphics[width =0.315\linewidth]{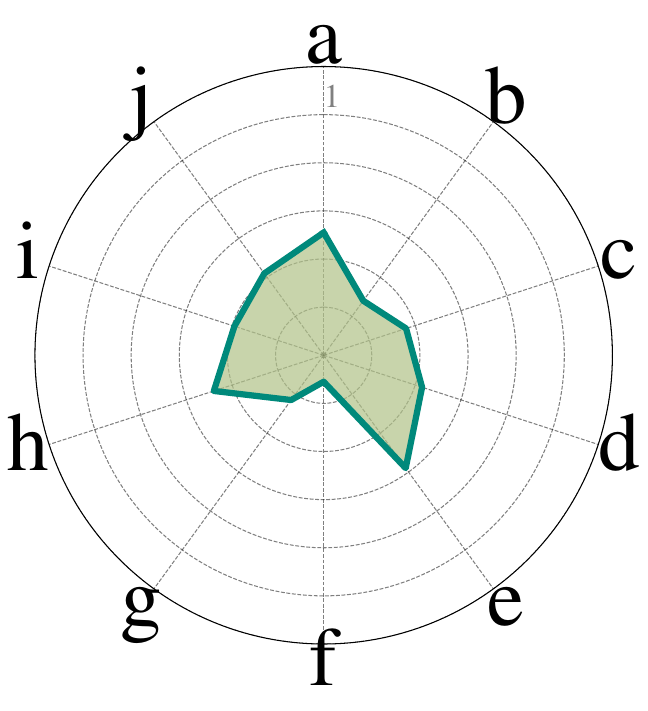}}
    \subfigure[\model~]{\includegraphics[width =0.315\linewidth]{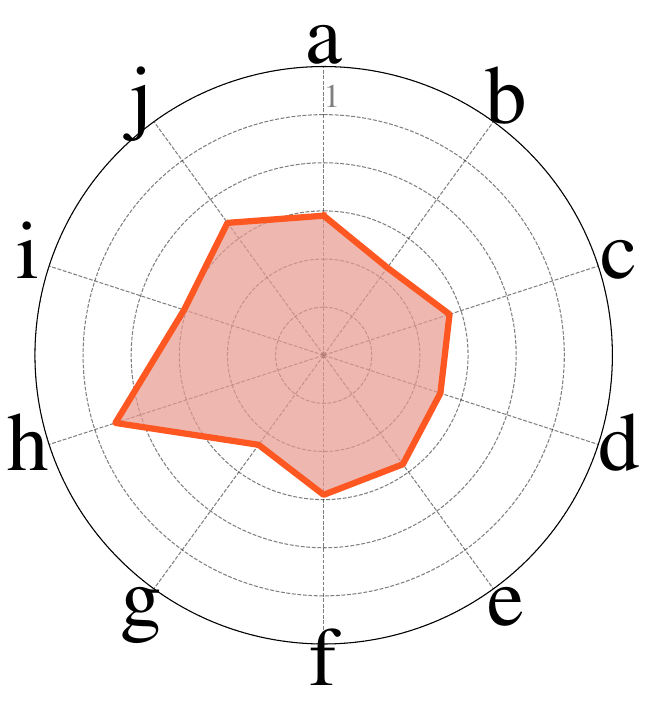}}
\caption{Performance magnitudes of w/o MoE,  MoE (Top-1) and \model~. a-j represent different downstream tasks.}
    \label{fig:eval_moe}
\end{figure}

\noindent \textbf{Effectiveness of Request-wise MoE Router.}
Figure~\ref{fig:eval_moe} illustrates the performance of different LoRA adapter fusion methods during runtime inference for dynamic end-user requests. \textit{w/o MoE} indicates direct averaging of all LoRA modules for all requests, which typically results in suboptimal performance due to the fact that not all stochastic multi-tasks benefit each other.  \textit{MoE (Top-1)} means based on the request to select the most similar LoRA for LLM inference. However, from Figure~\ref{fig:eval_moe}, we can see that \textit{w/o MoE}  Simply combining LoRAs results in worse performance because not all stochastic multi-tasks are beneficial to the others. Though \textit{MoE (Top-1)} signifies that the most similar LoRA is selected based on the request for LLM inference. Although this approach enhances overall performance, it can still be sub-optimal in practice, as even a single request may involve multiple tasks, and merely considering task differences is insufficient.
This observation highlights the critical importance of both the request-wise  MoE router in maintaining \model~ performance within diverse data and task demands.

\begin{figure}
    \centering
    \includegraphics[width=0.95\linewidth]{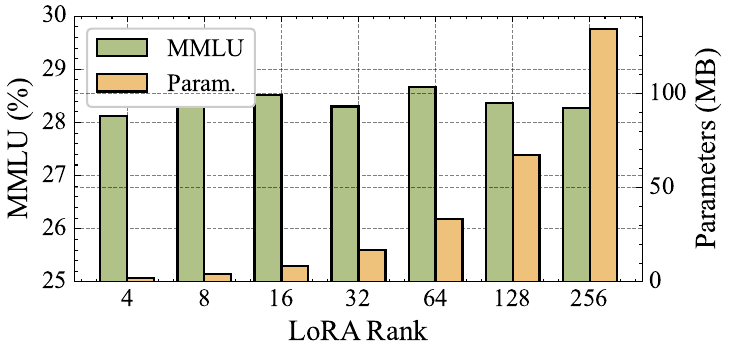}
    \caption{\new{MMLU performance and parameter size across LoRA ranks.}}
    \label{fig:diffrank}
\end{figure}

\noindent \new{\textbf{Impact of the Rank of LoRA Adapters.}
The rank $r$ controls the number of trainable parameters, where larger values of $r$ provide greater adapter capacity at the expense of increased training overhead. As illustrated in Figure~\ref{fig:diffrank}, performance initially improves with higher ranks but eventually saturates, offering diminishing returns despite the rapidly growing parameter size.}

\noindent \textbf{Overhead Analysis.} Offline tailoring utilizes a single-layer LSTM encoder-decoder and a dual-layer feed-forward network, adding minimal overhead and enhancing accuracy without affecting online inference. 
\new{During online inference, the soft MoE router introduces no additional trainable parameters and performs a single probabilistic routing computation.} \newnew{Implemented as a two-layer MLP with under 1K parameters, the DVFS controller imposes negligible overhead relative to billion-parameter LLMs.}
\new{Both modules first execute concurrently (< 10ms) with the prefill stage (> 100 ms; see Figure~\ref{fig:edge_TE}), then the DVFS decision for token $t+1$ is generated while token $t$ is being decoded, keeping the controller off the critical path. 
Meanwhile, the incremental energy cost of \model~ is on the milliwatt‑hour scale, rendering it negligible compared with
the watt‑hour‑level consumption of full‑model inference.}

\section{Related Work}
LLMs are highly memory-, compute-, and energy-intensive~\cite{GPT4, PaLM, llama, Llama_2, anthropic_claude, gemma}, driving most "billion-parameter" inference to the cloud~\cite{AIIndex2024, apple_ai, song2024powerinfer, yuan2023mobile, app_3, app5_robot, app6_laptops}. However, as edge devices become more powerful, there is increasing interest in executing LLM inference on the edge~\cite{apple_ai, song2024powerinfer, yuan2023mobile, app_3, app5_robot, app6_laptops, ExeGPT, Splitwise, PagedAttention, Just-in-time, llm_flash}, which enhances data privacy and enables real-time service delivery.
To address edge resource constraints, techniques such as model architecture search~\cite{jawahar2023llm, huang2024new, liu2024optimizing}, quantization~\cite{hubara2018quantized, GPT3.int8, Just-in-time, Deja, SmoothQuant}, pruning~\cite{cnn_pruning, Deja, Pruner-Zero, SparseGPT}, and knowledge distillation~\cite{DistiLLM, MiniLLM} have been proposed. However, most approaches focus solely on model-level optimization, neglecting system-level trade-offs like storage and weight efficiency. Advances in LLM compilers and software stacks~\cite{pytorch-1, tensorflow, deepspeed, huggingface_transformers} have enabled integration with co-processors and near-sensor processing~\cite{song2024powerinfer, ExeGPT, PagedAttention, Just-in-time, llm_flash, Splitwise, flexgen, Orca,zhao4,zhao5,zhao6,zhao7,zhao8}, but these often add computational and communication overhead, reducing edge device longevity and hindering concurrent application execution. Existing model customization methods~\cite{tambe2021edgebert, zhao2023approxcaliper, liberis2023differentiable, zhao2024felix, tam2024fedhybrid} cannot directly address task-agnostic LLMs, where generalization is crucial. The unique characteristics of LLM decoder layers and stochastic outputs present untapped opportunities for hardware optimization.
 For system-level optimization, DVFS~\cite{dvfsasplos, dfvs-4, dvfs-2, dvfs-3, bateni2020neuos} has been widely used to dynamically adjust processor voltage and frequency. However, most DVFS strategies are designed for discriminative models like CNNs and RNNs, treating networks as black boxes. Generative LLMs, with their auto-regressive inference and stochastic prompt variability, remain underexplored in this context.

\new{Edge–cloud collaboration offers a pragmatic middle ground between fully local and fully cloud‑based inference. In practice, the edge should remain the first line of execution for latency‑critical or privacy‑sensitive workloads, running compact SLMs that fit the device’s power and memory envelope. When a request exceeds local capacity—e.g., requires deeper reasoning, broader context, or larger knowledge—\model~ can transparently escalate the call to a cloud‑resident LLM. This selective offloading preserves real‑time responsiveness, keeps private data on‑device whenever possible, and amortizes bandwidth and compute costs by invoking the cloud only for the fraction of tasks that truly need it.}

\section{Conclusion}


\new{\model~ integrates offline generative pruning and online latency-aware optimization to enhance edge deployment of LLMs. Offline, it dynamically generates device-specific pruning configurations via a generative framework. Online, \model~ employs a layer-wise DVFS strategy to optimize energy efficiency within latency constraints, using an MoE router with multiple LoRA adapters to support diverse tasks. Supported by a dedicated 28nm hardware accelerator, including specialized units for rapid LoRA adapter switching and fine-grained DVFS control, benchmarks confirm that \model~ significantly improves inference speed and energy efficiency while maintaining robust task performance, ideal for latency-sensitive, energy-constrained edge scenarios.}

\section*{Acknowledgment}
\new{We sincerely thank the anonymous ATC'25 reviewers,
and our shepherd, Dr. Joel Wolfrath, for their insightful suggestions. This work is supported in part by the Science and Technology Development Fund of Macau (0107/2024/RIA2), Joint Science and Technology Research Project with Hong Kong and Macau in Key Areas of Nansha District's Science and Technology Plan (EF2024-00180-IOTSC) and the Multi-Year Research Grant of University of Macau (MYRG-GRG2023-00211-IOTSC-UMDF, MYRG-GRG2024-00180-IOTSC). Please ask Dr. Li Li (llili@um.edu.mo) for correspondence.}

\bibliographystyle{plain}
\bibliography{main}

\begin{thebibliography}{100}

\bibitem{rewind2024}
Rewind AI.
\newblock Rewind: A better way to search your digital life, 2024.

\bibitem{ADPLL_2}
Tutu Ajayi, Sumanth Kamineni, Yaswanth~K Cherivirala, Morteza Fayazi, Kyumin Kwon, Mehdi Saligane, Shourya Gupta, Chien-Hen Chen, Dennis Sylvester, David Blaauw, et~al.
\newblock An open-source framework for autonomous soc design with analog block generation.
\newblock In {\em 2020 IFIP/IEEE 28th International Conference on Very Large Scale Integration (VLSI-SOC)}, pages 141--146. IEEE, 2020.

\bibitem{alammar2019illustrated}
Jay Alammar.
\newblock The illustrated gpt-2 (visualizing transformer language models).
\newblock {\em Jalammar. github. io. https://jalammar. github. io/illustrated-gpt2}, 2019.

\bibitem{llm_flash}
Keivan Alizadeh, Iman Mirzadeh, Dmitry Belenko, Karen Khatamifard, Minsik Cho, Carlo C.~Del Mundo, Mohammad Rastegari, and Mehrdad Farajtabar.
\newblock {LLM} in a flash: Efficient large language model inference with limited memory.
\newblock {\em CoRR}, abs/2312.11514, 2023.

\bibitem{anthropic_claude}
Anthropic.
\newblock Claude: A family of ai models, 2024.

\bibitem{apple_ai}
Apple.
\newblock Apple intelligence.
\newblock \url{https://www.apple.com/apple-intelligence/}, 2024.

\bibitem{siri2024}
Apple.
\newblock Apple siri: Virtual assistant, 2024.

\bibitem{SliceGPT}
Saleh Ashkboos, Maximilian~L. Croci, Marcelo Gennari~Do Nascimento, Torsten Hoefler, and James Hensman.
\newblock Slicegpt: Compress large language models by deleting rows and columns.
\newblock {\em CoRR}, abs/2401.15024, 2024.

\bibitem{LDO_2}
Suyoung Bang, Wootaek Lim, Charles Augustine, Andres Malavasi, Muhammad Khellah, James Tschanz, and Vivek De.
\newblock 25.1 a fully synthesizable distributed and scalable all-digital ldo in 10nm cmos.
\newblock In {\em 2020 IEEE International Solid-State Circuits Conference-(ISSCC)}, pages 380--382. IEEE, 2020.

\bibitem{bateni2020neuos}
Soroush Bateni and Cong Liu.
\newblock $\{$NeuOS$\}$: A $\{$Latency-Predictable$\}$$\{$Multi-Dimensional$\}$ optimization framework for $\{$DNN-driven$\}$ autonomous systems.
\newblock In {\em 2020 USENIX Annual Technical Conference (USENIX ATC 20)}, pages 371--385, 2020.

\bibitem{output_train_1}
Emily~M. Bender, Timnit Gebru, Angelina McMillan{-}Major, and Shmargaret Shmitchell.
\newblock On the dangers of stochastic parrots: Can language models be too big?
\newblock In Madeleine~Clare Elish, William Isaac, and Richard~S. Zemel, editors, {\em FAccT '21: 2021 {ACM} Conference on Fairness, Accountability, and Transparency, Virtual Event / Toronto, Canada, March 3-10, 2021}, pages 610--623. {ACM}, 2021.

\bibitem{piqa}
Yonatan Bisk, Rowan Zellers, Ronan~Le Bras, Jianfeng Gao, and Yejin Choi.
\newblock {PIQA:} reasoning about physical commonsense in natural language.
\newblock In {\em The Thirty-Fourth {AAAI} Conference on Artificial Intelligence, {AAAI} 2020, The Thirty-Second Innovative Applications of Artificial Intelligence Conference, {IAAI} 2020, The Tenth {AAAI} Symposium on Educational Advances in Artificial Intelligence, {EAAI} 2020, New York, NY, USA, February 7-12, 2020}, pages 7432--7439. {AAAI} Press, 2020.

\bibitem{output_train_2}
Rishi Bommasani, Drew~A. Hudson, Ehsan Adeli, Russ~B. Altman, Simran Arora, Sydney von Arx, Michael~S. Bernstein, Jeannette Bohg, Antoine Bosselut, Emma Brunskill, Erik Brynjolfsson, Shyamal Buch, Dallas Card, Rodrigo Castellon, Niladri~S. Chatterji, Annie~S. Chen, Kathleen Creel, Jared~Quincy Davis, Dorottya Demszky, Chris Donahue, Moussa Doumbouya, Esin Durmus, Stefano Ermon, John Etchemendy, Kawin Ethayarajh, Li~Fei{-}Fei, Chelsea Finn, Trevor Gale, Lauren~E. Gillespie, Karan Goel, Noah~D. Goodman, Shelby Grossman, Neel Guha, Tatsunori Hashimoto, Peter Henderson, John Hewitt, Daniel~E. Ho, Jenny Hong, Kyle Hsu, Jing Huang, Thomas Icard, Saahil Jain, Dan Jurafsky, Pratyusha Kalluri, Siddharth Karamcheti, Geoff Keeling, Fereshte Khani, Omar Khattab, Pang~Wei Koh, Mark~S. Krass, Ranjay Krishna, Rohith Kuditipudi, and et~al.
\newblock On the opportunities and risks of foundation models.
\newblock {\em CoRR}, abs/2108.07258, 2021.

\bibitem{ppl-1}
Peter~F Brown, Stephen~A Della~Pietra, Vincent~J Della~Pietra, Jennifer~C Lai, and Robert~L Mercer.
\newblock An estimate of an upper bound for the entropy of english.
\newblock {\em Computational Linguistics}, 18(1):31--40, 1992.

\bibitem{input_unpredict}
Tom~B. Brown, Benjamin Mann, Nick Ryder, Melanie Subbiah, Jared Kaplan, Prafulla Dhariwal, Arvind Neelakantan, Pranav Shyam, Girish Sastry, Amanda Askell, Sandhini Agarwal, Ariel Herbert{-}Voss, Gretchen Krueger, Tom Henighan, Rewon Child, Aditya Ramesh, Daniel~M. Ziegler, Jeffrey Wu, Clemens Winter, Christopher Hesse, Mark Chen, Eric Sigler, Mateusz Litwin, Scott Gray, Benjamin Chess, Jack Clark, Christopher Berner, Sam McCandlish, Alec Radford, Ilya Sutskever, and Dario Amodei.
\newblock Language models are few-shot learners.
\newblock In Hugo Larochelle, Marc'Aurelio Ranzato, Raia Hadsell, Maria{-}Florina Balcan, and Hsuan{-}Tien Lin, editors, {\em Advances in Neural Information Processing Systems 33: Annual Conference on Neural Information Processing Systems 2020, NeurIPS 2020, December 6-12, 2020, virtual}, 2020.

\bibitem{LDO}
Chaitanya~K Chava and Jos{\'e} Silva-Mart{\'\i}nez.
\newblock A frequency compensation scheme for ldo voltage regulators.
\newblock {\em IEEE Transactions on Circuits and Systems I: Regular Papers}, 51(6):1041--1050, 2004.

\bibitem{bge-m3}
Jianlv Chen, Shitao Xiao, Peitian Zhang, Kun Luo, Defu Lian, and Zheng Liu.
\newblock Bge m3-embedding: Multi-lingual, multi-functionality, multi-granularity text embeddings through self-knowledge distillation, 2024.

\bibitem{Q_table}
Yonghun Choi, Seonghoon Park, and Hojung Cha.
\newblock Optimizing energy efficiency of browsers in energy-aware scheduling-enabled mobile devices.
\newblock In Stephen~A. Brewster, Geraldine Fitzpatrick, Anna~L. Cox, and Vassilis Kostakos, editors, {\em The 25th Annual International Conference on Mobile Computing and Networking, MobiCom 2019, Los Cabos, Mexico, October 21-25, 2019}, pages 48:1--48:16. {ACM}, 2019.

\bibitem{PaLM}
Aakanksha Chowdhery, Sharan Narang, Jacob Devlin, Maarten Bosma, Gaurav Mishra, Adam Roberts, Paul Barham, Hyung~Won Chung, Charles Sutton, Sebastian Gehrmann, Parker Schuh, Kensen Shi, Sasha Tsvyashchenko, Joshua Maynez, Abhishek Rao, Parker Barnes, Yi~Tay, Noam Shazeer, Vinodkumar Prabhakaran, Emily Reif, Nan Du, Ben Hutchinson, Reiner Pope, James Bradbury, Jacob Austin, Michael Isard, Guy Gur{-}Ari, Pengcheng Yin, Toju Duke, Anselm Levskaya, Sanjay Ghemawat, Sunipa Dev, Henryk Michalewski, Xavier Garcia, Vedant Misra, Kevin Robinson, Liam Fedus, Denny Zhou, Daphne Ippolito, David Luan, Hyeontaek Lim, Barret Zoph, Alexander Spiridonov, Ryan Sepassi, David Dohan, Shivani Agrawal, Mark Omernick, Andrew~M. Dai, Thanumalayan~Sankaranarayana Pillai, Marie Pellat, Aitor Lewkowycz, Erica Moreira, Rewon Child, Oleksandr Polozov, Katherine Lee, Zongwei Zhou, Xuezhi Wang, Brennan Saeta, Mark Diaz, Orhan Firat, Michele Catasta, Jason Wei, Kathy Meier{-}Hellstern, Douglas Eck, Jeff Dean, Slav Petrov, and Noah Fiedel.
\newblock Palm: Scaling language modeling with pathways.
\newblock {\em J. Mach. Learn. Res.}, 24:240:1--240:113, 2023.

\bibitem{ml-energy-leaderboard}
Jae-Won Chung, Jiachen Liu, Zhiyu Wu, Yuxuan Xia, and Mosharaf Chowdhury.
\newblock {ML.ENERGY} leaderboard.
\newblock \url{https://ml.energy/leaderboard}, 2023.

\bibitem{Scaling}
Aidan Clark, Diego de~Las~Casas, Aurelia Guy, Arthur Mensch, Michela Paganini, Jordan Hoffmann, Bogdan Damoc, Blake~A. Hechtman, Trevor Cai, Sebastian Borgeaud, George van~den Driessche, Eliza Rutherford, Tom Hennigan, Matthew~J. Johnson, Albin Cassirer, Chris Jones, Elena Buchatskaya, David Budden, Laurent Sifre, Simon Osindero, Oriol Vinyals, Marc'Aurelio Ranzato, Jack~W. Rae, Erich Elsen, Koray Kavukcuoglu, and Karen Simonyan.
\newblock Unified scaling laws for routed language models.
\newblock In Kamalika Chaudhuri, Stefanie Jegelka, Le~Song, Csaba Szepesv{\'{a}}ri, Gang Niu, and Sivan Sabato, editors, {\em International Conference on Machine Learning, {ICML} 2022, 17-23 July 2022, Baltimore, Maryland, {USA}}, volume 162 of {\em Proceedings of Machine Learning Research}, pages 4057--4086. {PMLR}, 2022.

\bibitem{boolq}
Christopher Clark, Kenton Lee, Ming{-}Wei Chang, Tom Kwiatkowski, Michael Collins, and Kristina Toutanova.
\newblock Boolq: Exploring the surprising difficulty of natural yes/no questions.
\newblock In Jill Burstein, Christy Doran, and Thamar Solorio, editors, {\em Proceedings of the 2019 Conference of the North American Chapter of the Association for Computational Linguistics: Human Language Technologies, {NAACL-HLT} 2019, Minneapolis, MN, USA, June 2-7, 2019, Volume 1 (Long and Short Papers)}, pages 2924--2936. Association for Computational Linguistics, 2019.

\bibitem{arc}
Peter Clark, Isaac Cowhey, Oren Etzioni, Tushar Khot, Ashish Sabharwal, Carissa Schoenick, and Oyvind Tafjord.
\newblock Think you have solved question answering? try arc, the ai2 reasoning challenge.
\newblock {\em arXiv:1803.05457v1}, 2018.

\bibitem{codecarbon}
codecarbon.
\newblock Track and reduce co2 emissions from your computing.
\newblock \url{https://codecarbon.io/}, 2023.

\bibitem{llama-dell}
Dell.
\newblock Llama 2: Inferencing on a single gpu.
\newblock \url{https://infohub.delltechnologies.com/zh-cn/t/llama-2-inferencing-on-a-single-gpu/}, 2023.

\bibitem{GPT3.int8}
Tim Dettmers, Mike Lewis, Younes Belkada, and Luke Zettlemoyer.
\newblock Llm.int8(): 8-bit matrix multiplication for transformers at scale.
\newblock In Sanmi Koyejo, S.~Mohamed, A.~Agarwal, Danielle Belgrave, K.~Cho, and A.~Oh, editors, {\em Advances in Neural Information Processing Systems 35: Annual Conference on Neural Information Processing Systems 2022, NeurIPS 2022, New Orleans, LA, USA, November 28 - December 9, 2022}, 2022.

\bibitem{Pruner-Zero}
Peijie Dong, Lujun Li, Zhenheng Tang, Xiang Liu, Xinglin Pan, Qiang Wang, and Xiaowen Chu.
\newblock Pruner-zero: Evolving symbolic pruning metric from scratch for large language models.
\newblock In {\em Forty-first International Conference on Machine Learning, {ICML} 2024, Vienna, Austria, July 21-27, 2024}. OpenReview.net, 2024.

\bibitem{User-aware}
Begum Egilmez, Matthew Schuchhardt, Gokhan Memik, Raid Ayoub, Niranjan Soundararajan, and Michael Kishinevsky.
\newblock User-aware frame rate management in android smartphones.
\newblock {\em {ACM} Trans. Embed. Comput. Syst.}, 16(5s):131:1--131:17, 2017.

\bibitem{OSDI_96}
Yasuhiro Endo, Zheng Wang, J.~Bradley Chen, and Margo~I. Seltzer.
\newblock Using latency to evaluate interactive system performance.
\newblock In Karin Petersen and Willy Zwaenepoel, editors, {\em Proceedings of the Second {USENIX} Symposium on Operating Systems Design and Implementation (OSDI), Seattle, Washington, USA, October 28-31, 1996}, pages 185--199. {ACM}, 1996.

\bibitem{LLM-Pruner}
Gongfan Fang, Xinyin Ma, Mingli Song, Michael~Bi Mi, and Xinchao Wang.
\newblock Depgraph: Towards any structural pruning.
\newblock In {\em {IEEE/CVF} Conference on Computer Vision and Pattern Recognition, {CVPR} 2023, Vancouver, BC, Canada, June 17-24, 2023}, pages 16091--16101. {IEEE}, 2023.

\bibitem{floridi2020gpt}
Luciano Floridi and Massimo Chiriatti.
\newblock Gpt-3: Its nature, scope, limits, and consequences.
\newblock {\em Minds and Machines}, 30:681--694, 2020.

\bibitem{SparseGPT}
Elias Frantar and Dan Alistarh.
\newblock Sparsegpt: Massive language models can be accurately pruned in one-shot.
\newblock In Andreas Krause, Emma Brunskill, Kyunghyun Cho, Barbara Engelhardt, Sivan Sabato, and Jonathan Scarlett, editors, {\em International Conference on Machine Learning, {ICML} 2023, 23-29 July 2023, Honolulu, Hawaii, {USA}}, volume 202 of {\em Proceedings of Machine Learning Research}, pages 10323--10337. {PMLR}, 2023.

\bibitem{freitag-al-onaizan-2017-beam}
Markus Freitag and Yaser Al{-}Onaizan.
\newblock Beam search strategies for neural machine translation.
\newblock In Thang Luong, Alexandra Birch, Graham Neubig, and Andrew~M. Finch, editors, {\em Proceedings of the First Workshop on Neural Machine Translation, NMT@ACL 2017, Vancouver, Canada, August 4, 2017}, pages 56--60. Association for Computational Linguistics, 2017.

\bibitem{app_3}
Daocheng Fu, Xin Li, Licheng Wen, Min Dou, Pinlong Cai, Botian Shi, and Yu~Qiao.
\newblock Drive like a human: Rethinking autonomous driving with large language models.
\newblock In {\em Proceedings of the IEEE/CVF Winter Conference on Applications of Computer Vision}, pages 910--919, 2024.

\bibitem{openlm2023openllama}
Xinyang Geng and Hao Liu.
\newblock Openllama: An open reproduction of llama.
\newblock \url{https://github.com/openlm-research/open_llama}, May 2023.

\bibitem{FLOPs_vgg}
Amir Gholami, Zhewei Yao, Sehoon Kim, Coleman Hooper, Michael~W. Mahoney, and Kurt Keutzer.
\newblock {AI} and memory wall.
\newblock {\em {IEEE} Micro}, 44(3):33--39, 2024.

\bibitem{github_copilot}
github.
\newblock {GitHub Copilot: Your AI pair programmer}.
\newblock \url{https://github.com/features/copilot}.

\bibitem{goodfellow2016deep}
Ian~J. Goodfellow, Yoshua Bengio, and Aaron~C. Courville.
\newblock {\em Deep Learning}.
\newblock Adaptive computation and machine learning. {MIT} Press, 2016.

\bibitem{googleAssistant2024}
Google.
\newblock Google assistant, 2024.

\bibitem{googleMLKit2024}
Google.
\newblock Ml kit smart reply, 2024.

\bibitem{MiniLLM}
Yuxian Gu, Li~Dong, Furu Wei, and Minlie Huang.
\newblock Minillm: Knowledge distillation of large language models.
\newblock In {\em The Twelfth International Conference on Learning Representations, {ICLR} 2024, Vienna, Austria, May 7-11, 2024}. OpenReview.net, 2024.

\bibitem{Mistify}
Peizhen Guo, Bo~Hu, and Wenjun Hu.
\newblock Mistify: Automating {DNN} model porting for on-device inference at the edge.
\newblock In James Mickens and Renata Teixeira, editors, {\em 18th {USENIX} Symposium on Networked Systems Design and Implementation, {NSDI} 2021, April 12-14, 2021}, pages 705--719. {USENIX} Association, 2021.

\bibitem{zhao8}
Pengyu He, Yuanzhe Zhao, Heng Xie, Yang Wang, Shouyi Yin, Li~Li, Yan Zhu, Rui~P Martins, Chi-Hang Chan, and Minglei Zhang.
\newblock A reconfigurable floating-point compute-in-memory with analog exponent pre-processes.
\newblock {\em IEEE Solid-State Circuits Letters}, 2024.

\bibitem{zhao2}
Pengyu He, Yuanzhe Zhao, Heng Xie, Yang Wang, Shouyi Yin, Li~Li, Yan Zhu, Rui~Paulo Martins, Chi-Hang Chan, and Minglei Zhang.
\newblock A 28nm 314.6 tlfops/w reconfigurable floating-point analog compute-in-memory macro with exponent approximation and two-stage sharing td-adc.
\newblock In {\em 2024 IEEE Custom Integrated Circuits Conference (CICC)}, pages 1--2. IEEE, 2024.

\bibitem{Channel_pruning}
Yihui He, Xiangyu Zhang, and Jian Sun.
\newblock Channel pruning for accelerating very deep neural networks.
\newblock In {\em {IEEE} International Conference on Computer Vision, {ICCV} 2017, Venice, Italy, October 22-29, 2017}, pages 1398--1406. {IEEE} Computer Society, 2017.

\bibitem{MMLU}
Dan Hendrycks, Collin Burns, Steven Basart, Andy Zou, Mantas Mazeika, Dawn Song, and Jacob Steinhardt.
\newblock Measuring massive multitask language understanding.
\newblock {\em Proceedings of the International Conference on Learning Representations (ICLR)}, 2021.

\bibitem{dvfs-2}
Sebastian Herbert and Diana Marculescu.
\newblock Variation-aware dynamic voltage/frequency scaling.
\newblock In {\em 15th International Conference on High-Performance Computer Architecture {(HPCA-15} 2009), 14-18 February 2009, Raleigh, North Carolina, {USA}}, pages 301--312. {IEEE} Computer Society, 2009.

\bibitem{dfvs-4}
Robert Hesse and Natalie D.~Enright Jerger.
\newblock Improving {DVFS} in nocs with coherence prediction.
\newblock In Andr{\'{e}} Ivanov, Diana Marculescu, Partha~Pratim Pande, Jos{\'{e}} Flich, and Karthik Pattabiraman, editors, {\em Proceedings of the 9th International Symposium on Networks-on-Chip, {NOCS} 2015, Vancouver, BC, Canada, September 28-30, 2015}, pages 24:1--24:8. {ACM}, 2015.

\bibitem{hochreiter1997long}
Sepp Hochreiter and J{\"{u}}rgen Schmidhuber.
\newblock Long short-term memory.
\newblock {\em Neural Comput.}, 9(8):1735--1780, 1997.

\bibitem{output_sampling}
Ari Holtzman, Jan Buys, Li~Du, Maxwell Forbes, and Yejin Choi.
\newblock The curious case of neural text degeneration.
\newblock In {\em 8th International Conference on Learning Representations, {ICLR} 2020, Addis Ababa, Ethiopia, April 26-30, 2020}. OpenReview.net, 2020.

\bibitem{ADPLL}
Terng-Yin Hsu, Bai-Jue Shieh, and Chen-Yi Lee.
\newblock An all-digital phase-locked loop (adpll)-based clock recovery circuit.
\newblock {\em IEEE Journal of Solid-State Circuits}, 34(8):1063--1073, 1999.

\bibitem{LoRA}
Edward~J. {Hu}, Yelong {Shen}, Phillip {Wallis}, Zeyuan {Allen-Zhu}, Yuanzhi {Li}, Shean {Wang}, Lu~{Wang}, and Weizhu {Chen}.
\newblock {LoRA: Low-Rank Adaptation of Large Language Models}.
\newblock {\em arXiv e-prints}, page arXiv:2106.09685, June 2021.

\bibitem{CAMA}
Yi~Huang, Zhiyu Chen, Dai Li, and Kaiyuan Yang.
\newblock {CAMA:} energy and memory efficient automata processing in content-addressable memories.
\newblock In {\em {IEEE} International Symposium on High-Performance Computer Architecture, {HPCA} 2022, Seoul, South Korea, April 2-6, 2022}, pages 25--37. {IEEE}, 2022.

\bibitem{huang2024new}
Yingbing Huang, Lily~Jiaxin Wan, Hanchen Ye, Manvi Jha, Jinghua Wang, Yuhong Li, Xiaofan Zhang, and Deming Chen.
\newblock New solutions on llm acceleration, optimization, and application.
\newblock In {\em Proceedings of the 61st ACM/IEEE Design Automation Conference}, pages 1--4, 2024.

\bibitem{hubara2018quantized}
Itay Hubara, Matthieu Courbariaux, Daniel Soudry, Ran El-Yaniv, and Yoshua Bengio.
\newblock Quantized neural networks: Training neural networks with low precision weights and activations.
\newblock {\em Journal of Machine Learning Research}, 18(187):1--30, 2018.

\bibitem{huggingface_transformers}
{HuggingFace Team}.
\newblock Transformers documentation, 2024.

\bibitem{Just-in-time}
Mohamed~Assem Ibrahim, Shaizeen Aga, Ada Li, Suchita Pati, and Mahzabeen Islam.
\newblock Just-in-time quantization with processing-in-memory for efficient ml training, 2023.

\bibitem{mem2024}
Mem Inc.
\newblock Mem: Your ai-powered assistant, 2024.

\bibitem{moe}
Robert~A. Jacobs, Michael~I. Jordan, Steven~J. Nowlan, and Geoffrey~E. Hinton.
\newblock Adaptive mixtures of local experts.
\newblock {\em Neural Comput.}, 3(1):79--87, 1991.

\bibitem{jawahar2023llm}
Ganesh Jawahar, Muhammad Abdul-Mageed, Laks~VS Lakshmanan, and Dujian Ding.
\newblock Llm performance predictors are good initializers for architecture search.
\newblock {\em arXiv preprint arXiv:2310.16712}, 2023.

\bibitem{kabilankb2024llama}
Kabilankb.
\newblock Robot control using llama: Bridging ai and robotics.
\newblock \href{https://medium.com/@kabilankb2003/robot-control-using-llama-bridging-ai-and-robotics-4bf8b37b953c}{Medium}, 2024.
\newblock Oct 13, 2024.

\bibitem{RL_1}
Leslie~Pack Kaelbling, Michael~L Littman, and Andrew~W Moore.
\newblock Reinforcement learning: A survey.
\newblock {\em Journal of artificial intelligence research}, 4:237--285, 1996.

\bibitem{Scaling_up}
Jared Kaplan, Sam McCandlish, Tom Henighan, Tom~B. Brown, Benjamin Chess, Rewon Child, Scott Gray, Alec Radford, Jeffrey Wu, and Dario Amodei.
\newblock Scaling laws for neural language models.
\newblock {\em CoRR}, abs/2001.08361, 2020.

\bibitem{Micro_LLM}
Byeongho Kim, Sanghoon Cha, Sangsoo Park, Jieun Lee, Sukhan Lee, Shinhaeng Kang, Jinin So, Kyungsoo Kim, Jin Jung, Jong{-}Geon Lee, Sunjung Lee, Yoonah Paik, Hyeonsu Kim, Jin{-}Seong Kim, Won{-}Jo Lee, Yuhwan Ro, Yeongon Cho, Jin~Hyun Kim, Joon{-}Ho Song, Jaehoon Yu, Seungwon Lee, Jeonghyeon Cho, and Kyomin Sohn.
\newblock The breakthrough memory solutions for improved performance on {LLM} inference.
\newblock {\em {IEEE} Micro}, 44(3):40--48, 2024.

\bibitem{dvfs-3}
Wonyoung Kim, Meeta~Sharma Gupta, Gu{-}Yeon Wei, and David~M. Brooks.
\newblock System level analysis of fast, per-core {DVFS} using on-chip switching regulators.
\newblock In {\em 14th International Conference on High-Performance Computer Architecture {(HPCA-14} 2008), 16-20 February 2008, Salt Lake City, UT, {USA}}, pages 123--134. {IEEE} Computer Society, 2008.

\bibitem{edge_energy}
Young~Geun Kim, Minyong Kim, and Sung~Woo Chung.
\newblock Enhancing energy efficiency of multimedia applications in heterogeneous mobile multi-core processors.
\newblock {\em {IEEE} Trans. Computers}, 66(11):1878--1889, 2017.

\bibitem{power_2}
Young~Geun Kim, Minyong Kim, Jae~Min Kim, Minyoung Sung, and Sung~Woo Chung.
\newblock A novel gpu power model for accurate smartphone power breakdown.
\newblock {\em ETRI journal}, 37(1):157--164, 2015.

\bibitem{energy_survey}
Young~Geun Kim, Joonho Kong, and Sung~Woo Chung.
\newblock A survey on recent os-level energy management techniques for mobile processing units.
\newblock {\em {IEEE} Trans. Parallel Distributed Syst.}, 29(10):2388--2401, 2018.

\bibitem{autoscale}
Young~Geun Kim and Carole{-}Jean Wu.
\newblock Autoscale: Energy efficiency optimization for stochastic edge inference using reinforcement learning.
\newblock In {\em 53rd Annual {IEEE/ACM} International Symposium on Microarchitecture, {MICRO} 2020, Athens, Greece, October 17-21, 2020}, pages 1082--1096. {IEEE}, 2020.

\bibitem{power_1}
Youngsok Kim, Joonsung Kim, Dongju Chae, Daehyun Kim, and Jangwoo Kim.
\newblock $\mu$layer: Low latency on-device inference using cooperative single-layer acceleration and processor-friendly quantization.
\newblock In {\em Proceedings of the Fourteenth EuroSys Conference 2019}, pages 1--15, 2019.

\bibitem{DistiLLM}
Jongwoo Ko, Sungnyun Kim, Tianyi Chen, and Se{-}Young Yun.
\newblock Distillm: Towards streamlined distillation for large language models.
\newblock In {\em Forty-first International Conference on Machine Learning, {ICML} 2024, Vienna, Austria, July 21-27, 2024}. OpenReview.net, 2024.

\bibitem{koonce2021resnet}
Brett Koonce and Brett Koonce.
\newblock Resnet 50.
\newblock {\em Convolutional neural networks with swift for tensorflow: image recognition and dataset categorization}, pages 63--72, 2021.

\bibitem{eNVM_2}
Srivatsan Krishnan, Zishen Wan, Kshitij Bhardwaj, Paul Whatmough, Aleksandra Faust, Sabrina Neuman, Gu-Yeon Wei, David Brooks, and Vijay~Janapa Reddi.
\newblock Automatic domain-specific soc design for autonomous unmanned aerial vehicles.
\newblock In {\em 2022 55th IEEE/ACM International Symposium on Microarchitecture (MICRO)}, pages 300--317. IEEE, 2022.

\bibitem{PagedAttention}
Woosuk Kwon, Zhuohan Li, Siyuan Zhuang, Ying Sheng, Lianmin Zheng, Cody~Hao Yu, Joseph Gonzalez, Hao Zhang, and Ion Stoica.
\newblock Efficient memory management for large language model serving with pagedattention.
\newblock In Jason Flinn, Margo~I. Seltzer, Peter Druschel, Antoine Kaufmann, and Jonathan Mace, editors, {\em Proceedings of the 29th Symposium on Operating Systems Principles, {SOSP} 2023, Koblenz, Germany, October 23-26, 2023}, pages 611--626. {ACM}, 2023.

\bibitem{sosp}
Woosuk Kwon, Zhuohan Li, Siyuan Zhuang, Ying Sheng, Lianmin Zheng, Cody~Hao Yu, Joseph Gonzalez, Hao Zhang, and Ion Stoica.
\newblock Efficient memory management for large language model serving with pagedattention.
\newblock In Jason Flinn, Margo~I. Seltzer, Peter Druschel, Antoine Kaufmann, and Jonathan Mace, editors, {\em Proceedings of the 29th Symposium on Operating Systems Principles, {SOSP} 2023, Koblenz, Germany, October 23-26, 2023}, pages 611--626. {ACM}, 2023.

\bibitem{lecun2015deep}
Yann LeCun, Yoshua Bengio, and Geoffrey Hinton.
\newblock Deep learning.
\newblock {\em nature}, 521(7553):436--444, 2015.

\bibitem{eNVM_3}
Haitong Li, Mudit Bhargava, Paul~N Whatmough, and H-S~Philip Wong.
\newblock On-chip memory technology design space explorations for mobile deep neural network accelerators.
\newblock In {\em Proceedings of the 56th Annual Design Automation Conference 2019}, pages 1--6, 2019.

\bibitem{web_app_2}
Li~Li, Xiaorui Wang, and Feng Qin.
\newblock Energydx: Diagnosing energy anomaly in mobile apps by identifying the manifestation point.
\newblock In {\em 2020 IEEE 40th International Conference on Distributed Computing Systems (ICDCS)}, pages 256--266. IEEE, 2020.

\bibitem{RL_4}
Yuxi Li.
\newblock Deep reinforcement learning: An overview.
\newblock {\em arXiv preprint arXiv:1701.07274}, 2017.

\bibitem{li202512surveyreasoning}
Zhong-Zhi Li, Duzhen Zhang, Ming-Liang Zhang, Jiaxin Zhang, Zengyan Liu, Yuxuan Yao, Haotian Xu, Junhao Zheng, Pei-Jie Wang, Xiuyi Chen, Yingying Zhang, Fei Yin, Jiahua Dong, Zhiwei Li, Bao-Long Bi, Ling-Rui Mei, Junfeng Fang, Zhijiang Guo, Le~Song, and Cheng-Lin Liu.
\newblock From system 1 to system 2: A survey of reasoning large language models, 2025.

\bibitem{app5_robot}
Haicheng Liao, Zhenning Li, Huanming Shen, Wenxuan Zeng, Dongping Liao, Guofa Li, Shengbo~Eben Li, and Chengzhong Xu.
\newblock {BAT:} behavior-aware human-like trajectory prediction for autonomous driving.
\newblock {\em CoRR}, abs/2312.06371, 2023.

\bibitem{liberis2023differentiable}
Edgar Liberis and Nicholas~D Lane.
\newblock Differentiable neural network pruning to enable smart applications on microcontrollers.
\newblock {\em Proceedings of the ACM on Interactive, Mobile, Wearable and Ubiquitous Technologies}, 6(4):1--19, 2023.

\bibitem{application_3}
Chaofan Lin, Zhenhua Han, Chengruidong Zhang, Yuqing Yang, Fan Yang, Chen Chen, and Lili Qiu.
\newblock Parrot: Efficient serving of llm-based applications with semantic variable.
\newblock In Ada Gavrilovska and Douglas~B. Terry, editors, {\em 18th {USENIX} Symposium on Operating Systems Design and Implementation, {OSDI} 2024, Santa Clara, CA, USA, July 10-12, 2024}, pages 929--945. {USENIX} Association, 2024.

\bibitem{AWQ}
Ji~Lin, Jiaming Tang, Haotian Tang, Shang Yang, Xingyu Dang, and Song Han.
\newblock {AWQ:} activation-aware weight quantization for {LLM} compression and acceleration.
\newblock {\em CoRR}, abs/2306.00978, 2023.

\bibitem{RL_2}
Xue Lin, Yanzhi Wang, and Massoud Pedram.
\newblock A reinforcement learning-based power management framework for green computing data centers.
\newblock In {\em 2016 {IEEE} International Conference on Cloud Engineering, {IC2E} 2016, Berlin, Germany, April 4-8, 2016}, pages 135--138. {IEEE} Computer Society, 2016.

\bibitem{zhao1}
Jiahao Liu, Yuanzhe Zhao, Yan Zhu, Chi{-}Hang Chan, and Rui~Paulo Martins.
\newblock A weak puf-assisted strong {PUF} with inherent immunity to modeling attacks and ultra-low {BER}.
\newblock {\em {IEEE} Trans. Circuits Syst. {I} Regul. Pap.}, 69(12):4898--4907, 2022.

\bibitem{prompt_formats}
Pengfei Liu, Weizhe Yuan, Jinlan Fu, Zhengbao Jiang, Hiroaki Hayashi, and Graham Neubig.
\newblock Pre-train, prompt, and predict: {A} systematic survey of prompting methods in natural language processing.
\newblock {\em {ACM} Comput. Surv.}, 55(9):195:1--195:35, 2023.

\bibitem{liu2024optimizing}
Shu Liu, Asim Biswal, Audrey Cheng, Xiangxi Mo, Shiyi Cao, Joseph~E Gonzalez, Ion Stoica, and Matei Zaharia.
\newblock Optimizing llm queries in relational workloads.
\newblock {\em arXiv preprint arXiv:2403.05821}, 2024.

\bibitem{P-Tuning}
Xiao Liu, Kaixuan Ji, Yicheng Fu, Zhengxiao Du, Zhilin Yang, and Jie Tang.
\newblock P-tuning v2: Prompt tuning can be comparable to fine-tuning universally across scales and tasks.
\newblock {\em CoRR}, abs/2110.07602, 2021.

\bibitem{eNVM}
Zhi-Gang Liu, Paul~N Whatmough, Yuhao Zhu, and Matthew Mattina.
\newblock S2ta: Exploiting structured sparsity for energy-efficient mobile cnn acceleration.
\newblock In {\em 2022 IEEE International Symposium on High-Performance Computer Architecture (HPCA)}, pages 573--586. IEEE, 2022.

\bibitem{Deja}
Zichang Liu, Jue Wang, Tri Dao, Tianyi Zhou, Binhang Yuan, Zhao Song, Anshumali Shrivastava, Ce~Zhang, Yuandong Tian, Christopher R{\'{e}}, and Beidi Chen.
\newblock Deja vu: Contextual sparsity for efficient llms at inference time.
\newblock In Andreas Krause, Emma Brunskill, Kyunghyun Cho, Barbara Engelhardt, Sivan Sabato, and Jonathan Scarlett, editors, {\em International Conference on Machine Learning, {ICML} 2023, 23-29 July 2023, Honolulu, Hawaii, {USA}}, volume 202 of {\em Proceedings of Machine Learning Research}, pages 22137--22176. {PMLR}, 2023.

\bibitem{micro_15}
Daniel Lo, Taejoon Song, and G~Edward Suh.
\newblock Prediction-guided performance-energy trade-off for interactive applications.
\newblock In {\em Proceedings of the 48th International Symposium on Microarchitecture}, pages 508--520, 2015.

\bibitem{Ptb}
Mitchell~P. Marcus, Beatrice Santorini, and Mary~Ann Marcinkiewicz.
\newblock Building a large annotated corpus of english: The penn treebank.
\newblock {\em Comput. Linguistics}, 19(2):313--330, 1993.

\bibitem{ShortGPT}
Xin Men, Mingyu Xu, Qingyu Zhang, Bingning Wang, Hongyu Lin, Yaojie Lu, Xianpei Han, and Weipeng Chen.
\newblock Shortgpt: Layers in large language models are more redundant than you expect.
\newblock {\em CoRR}, abs/2403.03853, 2024.

\bibitem{Wikitext2}
Stephen Merity, Caiming Xiong, James Bradbury, and Richard Socher.
\newblock Pointer sentinel mixture models.
\newblock In {\em 5th International Conference on Learning Representations, {ICLR} 2017, Toulon, France, April 24-26, 2017, Conference Track Proceedings}. OpenReview.net, 2017.

\bibitem{gemma}
Thomas Mesnard, Cassidy Hardin, Robert Dadashi, Surya Bhupatiraju, Shreya Pathak, Laurent Sifre, Morgane Rivi{\`{e}}re, Mihir~Sanjay Kale, Juliette Love, Pouya Tafti, L{\'{e}}onard Hussenot, Aakanksha Chowdhery, Adam Roberts, Aditya Barua, Alex Botev, Alex Castro{-}Ros, Ambrose Slone, Am{\'{e}}lie H{\'{e}}liou, Andrea Tacchetti, Anna Bulanova, Antonia Paterson, Beth Tsai, Bobak Shahriari, Charline~Le Lan, Christopher~A. Choquette{-}Choo, Cl{\'{e}}ment Crepy, Daniel Cer, Daphne Ippolito, David Reid, Elena Buchatskaya, Eric Ni, Eric Noland, Geng Yan, George Tucker, George{-}Christian Muraru, Grigory Rozhdestvenskiy, Henryk Michalewski, Ian Tenney, Ivan Grishchenko, Jacob Austin, James Keeling, Jane Labanowski, Jean{-}Baptiste Lespiau, Jeff Stanway, Jenny Brennan, Jeremy Chen, Johan Ferret, Justin Chiu, and et~al.
\newblock Gemma: Open models based on gemini research and technology.
\newblock {\em CoRR}, abs/2403.08295, 2024.

\bibitem{microsoft_bing_new_features}
microsoft.
\newblock {Your Everyday AI Companion | Microsoft Bing}.
\newblock \url{https://www.bing.com/new}.

\bibitem{azureCognitive2024}
Microsoft.
\newblock Azure cognitive services - text analytics: Smart reply, 2024.

\bibitem{deepspeed}
{Microsoft DeepSpeed Team}.
\newblock Deepspeed: Advancing the science of ai through efficient training of large models, 2024.

\bibitem{OpenbookQA}
Todor Mihaylov, Peter Clark, Tushar Khot, and Ashish Sabharwal.
\newblock Can a suit of armor conduct electricity? {A} new dataset for open book question answering.
\newblock In Ellen Riloff, David Chiang, Julia Hockenmaier, and Jun'ichi Tsujii, editors, {\em Proceedings of the 2018 Conference on Empirical Methods in Natural Language Processing, Brussels, Belgium, October 31 - November 4, 2018}, pages 2381--2391. Association for Computational Linguistics, 2018.

\bibitem{RL_3}
Volodymyr Mnih, Koray Kavukcuoglu, David Silver, Andrei~A Rusu, Joel Veness, Marc~G Bellemare, Alex Graves, Martin Riedmiller, Andreas~K Fidjeland, Georg Ostrovski, et~al.
\newblock Human-level control through deep reinforcement learning.
\newblock {\em nature}, 518(7540):529--533, 2015.

\bibitem{importance_pruning}
Pavlo Molchanov, Arun Mallya, Stephen Tyree, Iuri Frosio, and Jan Kautz.
\newblock Importance estimation for neural network pruning.
\newblock In {\em {IEEE} Conference on Computer Vision and Pattern Recognition, {CVPR} 2019, Long Beach, CA, USA, June 16-20, 2019}, pages 11264--11272. Computer Vision Foundation / {IEEE}, 2019.

\bibitem{cnn_pruning}
Pavlo Molchanov, Stephen Tyree, Tero Karras, Timo Aila, and Jan Kautz.
\newblock Pruning convolutional neural networks for resource efficient inference.
\newblock {\em arXiv preprint arXiv:1611.06440}, 2016.

\bibitem{nano}
nano.
\newblock Nvidia jetson nano, 2023.

\bibitem{Twig}
Rajiv Nishtala, Vinicius Petrucci, Paul~M. Carpenter, and Magnus Sj{\"{a}}lander.
\newblock Twig: Multi-agent task management for colocated latency-critical cloud services.
\newblock In {\em {IEEE} International Symposium on High Performance Computer Architecture, {HPCA} 2020, San Diego, CA, USA, February 22-26, 2020}, pages 167--179. {IEEE}, 2020.

\bibitem{nvidia_jetson_orin}
NVIDIA.
\newblock Nvidia jetson orin - autonomous machines - nvidia.
\newblock \url{https://www.nvidia.com/en-us/autonomous-machines/embedded-systems/jetson-orin/}, 2024.

\bibitem{ExeGPT}
Hyungjun Oh, Kihong Kim, Jaemin Kim, Sungkyun Kim, Junyeol Lee, Du{-}seong Chang, and Jiwon Seo.
\newblock Exegpt: Constraint-aware resource scheduling for {LLM} inference.
\newblock In Rajiv Gupta, Nael~B. Abu{-}Ghazaleh, Madan Musuvathi, and Dan Tsafrir, editors, {\em Proceedings of the 29th {ACM} International Conference on Architectural Support for Programming Languages and Operating Systems, Volume 2, {ASPLOS} 2024, La Jolla, CA, USA, 27 April 2024- 1 May 2024}, pages 369--384. {ACM}, 2024.

\bibitem{chatgpt}
OpenAI.
\newblock Chatgpt, 2022.

\bibitem{GPT4}
OpenAI.
\newblock {GPT-4} technical report.
\newblock {\em CoRR}, abs/2303.08774, 2023.

\bibitem{otter2024}
Otter.ai.
\newblock Otter: Transcription and note-taking with ai, 2024.

\bibitem{Splitwise}
Pratyush Patel, Esha Choukse, Chaojie Zhang, {\'{I}}{\~{n}}igo Goiri, Aashaka Shah, Saeed Maleki, and Ricardo Bianchini.
\newblock Splitwise: Efficient generative {LLM} inference using phase splitting.
\newblock {\em CoRR}, abs/2311.18677, 2023.

\bibitem{dvfsasplos}
Leonardo Piga, Iyswarya Narayanan, Aditya Sundarrajan, Matt Skach, Qingyuan Deng, Biswadip Maity, Manoj Chakkaravarthy, Alison Huang, Abhishek Dhanotia, and Parth Malani.
\newblock Expanding datacenter capacity with {DVFS} boosting: {A} safe and scalable deployment experience.
\newblock In Rajiv Gupta, Nael~B. Abu{-}Ghazaleh, Madan Musuvathi, and Dan Tsafrir, editors, {\em Proceedings of the 29th {ACM} International Conference on Architectural Support for Programming Languages and Operating Systems, Volume 1, {ASPLOS} 2024, La Jolla, CA, USA, 27 April 2024- 1 May 2024}, pages 150--165. {ACM}, 2024.

\bibitem{pytorch-1}
{PyTorch Contributors}.
\newblock Pytorch: An open source machine learning framework, 2024.

\bibitem{qualcomm2024llama}
{Qualcomm Technologies, Inc.}
\newblock {Llama-v2-7B-Chat Quantized Model}.
\newblock \url{https://aihub.qualcomm.com/models/llama_v2_7b_chat_quantized}, 2024.
\newblock Qualcomm AI Hub, 2024.

\bibitem{output_token}
Alec Radford, Jeffrey Wu, Rewon Child, David Luan, Dario Amodei, Ilya Sutskever, et~al.
\newblock Language models are unsupervised multitask learners.
\newblock {\em OpenAI blog}, 1(8):9, 2019.

\bibitem{WinoGrande}
Keisuke Sakaguchi, Ronan~Le Bras, Chandra Bhagavatula, and Yejin Choi.
\newblock Winogrande: An adversarial winograd schema challenge at scale.
\newblock {\em arXiv preprint arXiv:1907.10641}, 2019.

\bibitem{ShazeerMMDLHD17}
Noam Shazeer, Azalia Mirhoseini, Krzysztof Maziarz, Andy Davis, Quoc~V. Le, Geoffrey~E. Hinton, and Jeff Dean.
\newblock Outrageously large neural networks: The sparsely-gated mixture-of-experts layer.
\newblock In {\em 5th International Conference on Learning Representations, {ICLR} 2017, Toulon, France, April 24-26, 2017, Conference Track Proceedings}. OpenReview.net, 2017.

\bibitem{flexgen}
Ying Sheng, Lianmin Zheng, Binhang Yuan, Zhuohan Li, Max Ryabinin, Beidi Chen, Percy Liang, Christopher R{\'e}, Ion Stoica, and Ce~Zhang.
\newblock Flexgen: High-throughput generative inference of large language models with a single gpu.
\newblock In {\em International Conference on Machine Learning}, pages 31094--31116. PMLR, 2023.

\bibitem{app4_robot}
Ishika Singh, Valts Blukis, Arsalan Mousavian, Ankit Goyal, Danfei Xu, Jonathan Tremblay, Dieter Fox, Jesse Thomason, and Animesh Garg.
\newblock Progprompt: Generating situated robot task plans using large language models.
\newblock In {\em {IEEE} International Conference on Robotics and Automation, {ICRA} 2023, London, UK, May 29 - June 2, 2023}, pages 11523--11530, London, UK, 2023. {IEEE}.

\bibitem{app2_clinical}
Karan Singhal, Shekoofeh Azizi, Tao Tu, S.~Sara Mahdavi, Jason Wei, Hyung~Won Chung, Nathan Scales, Ajay~Kumar Tanwani, Heather Cole{-}Lewis, Stephen Pfohl, Perry Payne, Martin Seneviratne, Paul Gamble, Chris Kelly, Nathaneal Sch{\"{a}}rli, Aakanksha Chowdhery, Philip~Andrew Mansfield, Blaise~Ag{\"{u}}era y~Arcas, Dale~R. Webster, Gregory~S. Corrado, Yossi Matias, Katherine Chou, Juraj Gottweis, Nenad Tomasev, Yun Liu, Alvin Rajkomar, Joelle~K. Barral, Christopher Semturs, Alan Karthikesalingam, and Vivek Natarajan.
\newblock Large language models encode clinical knowledge.
\newblock {\em CoRR}, abs/2212.13138, 2022.

\bibitem{song2023llm}
Chan~Hee Song, Jiaman Wu, Clayton Washington, Brian~M Sadler, Wei-Lun Chao, and Yu~Su.
\newblock Llm-planner: Few-shot grounded planning for embodied agents with large language models.
\newblock In {\em Proceedings of the IEEE/CVF International Conference on Computer Vision}, pages 2998--3009, 2023.

\bibitem{song2024powerinfer}
Yixin Song, Zeyu Mi, Haotong Xie, and Haibo Chen.
\newblock Powerinfer: Fast large language model serving with a consumer-grade gpu.
\newblock In {\em Proceedings of the ACM SIGOPS 30th Symposium on Operating Systems Principles}, pages 590--606, 2024.

\bibitem{bbh}
Aarohi Srivastava, Abhinav Rastogi, Abhishek Rao, Abu Awal~Md Shoeb, Abubakar Abid, Adam Fisch, Adam~R. Brown, Adam Santoro, Aditya Gupta, Adri{\`{a}} Garriga{-}Alonso, Agnieszka Kluska, Aitor Lewkowycz, Akshat Agarwal, Alethea Power, Alex Ray, Alex Warstadt, Alexander~W. Kocurek, Ali Safaya, Ali Tazarv, Alice Xiang, Alicia Parrish, Allen Nie, Aman Hussain, Amanda Askell, Amanda Dsouza, Ameet Rahane, Anantharaman~S. Iyer, Anders Andreassen, Andrea Santilli, Andreas Stuhlm{\"{u}}ller, Andrew~M. Dai, Andrew La, Andrew~K. Lampinen, Andy Zou, Angela Jiang, Angelica Chen, Anh Vuong, Animesh Gupta, Anna Gottardi, Antonio Norelli, Anu Venkatesh, Arash Gholamidavoodi, Arfa Tabassum, Arul Menezes, Arun Kirubarajan, Asher Mullokandov, Ashish Sabharwal, Austin Herrick, Avia Efrat, Aykut Erdem, Ayla Karakas, and et~al.
\newblock Beyond the imitation game: Quantifying and extrapolating the capabilities of language models.
\newblock {\em CoRR}, abs/2206.04615, 2022.

\bibitem{AIIndex2024}
stanford.
\newblock Ai index report.
\newblock \url{https://aiindex.stanford.edu/report/}, 2024.
\newblock Accessed: 2024-07.

\bibitem{RAM}
{Statista Inc.}
\newblock Mobile ram usage worldwide from 1q-19 to 1q-21 (in gb per device).
\newblock www.statista.com/statistics/1057679/mobile-ram-usage-worldwide-by-average-size-per-device/, 2021.

\bibitem{tam2024fedhybrid}
Kahou Tam, Chunlin Tian, Li~Li, Haikai Zhao, and ChengZhong Xu.
\newblock Fedhybrid: Breaking the memory wall of federated learning via hybrid tensor management.
\newblock In {\em Proceedings of the 22nd ACM Conference on Embedded Networked Sensor Systems}, pages 394--408, 2024.

\bibitem{tambe2021edgebert}
Thierry Tambe, Coleman Hooper, Lillian Pentecost, Tianyu Jia, En-Yu Yang, Marco Donato, Victor Sanh, Paul Whatmough, Alexander~M Rush, David Brooks, et~al.
\newblock Edgebert: Sentence-level energy optimizations for latency-aware multi-task nlp inference.
\newblock In {\em MICRO-54: 54th Annual IEEE/ACM International Symposium on Microarchitecture}, pages 830--844, 2021.

\bibitem{tensorflow}
{TensorFlow Contributors}.
\newblock Tensorflow: An open source machine learning framework for everyone, 2024.

\bibitem{app1-medicine}
Arun~James Thirunavukarasu, Darren Shu~Jeng Ting, Kabilan Elangovan, Laura Gutierrez, Ting~Fang Tan, and Daniel Shu~Wei Ting.
\newblock Large language models in medicine.
\newblock {\em Nature medicine}, 29(8):1930--1940, 2023.

\bibitem{tian2022harmony}
Chunlin Tian, Li~Li, Zhan Shi, Jun Wang, and ChengZhong Xu.
\newblock Harmony: Heterogeneity-aware hierarchical management for federated learning system.
\newblock In {\em 2022 55th IEEE/ACM International Symposium on Microarchitecture (MICRO)}, pages 631--645. IEEE, 2022.

\bibitem{tian2024hydralora}
Chunlin Tian, Zhan Shi, Zhijiang Guo, Li~Li, and Cheng-Zhong Xu.
\newblock Hydralora: An asymmetric lora architecture for efficient fine-tuning.
\newblock {\em Advances in Neural Information Processing Systems}, 37:9565--9584, 2024.

\bibitem{llama}
Hugo Touvron, Thibaut Lavril, Gautier Izacard, Xavier Martinet, Marie{-}Anne Lachaux, Timoth{\'{e}}e Lacroix, Baptiste Rozi{\`{e}}re, Naman Goyal, Eric Hambro, Faisal Azhar, Aur{\'{e}}lien Rodriguez, Armand Joulin, Edouard Grave, and Guillaume Lample.
\newblock Llama: Open and efficient foundation language models.
\newblock {\em CoRR}, abs/2302.13971, 2023.

\bibitem{Llama_2}
Hugo Touvron, Louis Martin, Kevin Stone, Peter Albert, Amjad Almahairi, Yasmine Babaei, Nikolay Bashlykov, Soumya Batra, Prajjwal Bhargava, Shruti Bhosale, Dan Bikel, Lukas Blecher, Cristian Canton{-}Ferrer, Moya Chen, Guillem Cucurull, David Esiobu, Jude Fernandes, Jeremy Fu, Wenyin Fu, Brian Fuller, Cynthia Gao, Vedanuj Goswami, Naman Goyal, Anthony Hartshorn, Saghar Hosseini, Rui Hou, Hakan Inan, Marcin Kardas, Viktor Kerkez, Madian Khabsa, Isabel Kloumann, Artem Korenev, Punit~Singh Koura, Marie{-}Anne Lachaux, Thibaut Lavril, Jenya Lee, Diana Liskovich, Yinghai Lu, Yuning Mao, Xavier Martinet, Todor Mihaylov, Pushkar Mishra, Igor Molybog, Yixin Nie, Andrew Poulton, Jeremy Reizenstein, Rashi Rungta, Kalyan Saladi, Alan Schelten, Ruan Silva, Eric~Michael Smith, Ranjan Subramanian, Xiaoqing~Ellen Tan, Binh Tang, Ross Taylor, Adina Williams, Jian~Xiang Kuan, Puxin Xu, Zheng Yan, Iliyan Zarov, Yuchen Zhang, Angela Fan, Melanie Kambadur, Sharan Narang, Aur{\'{e}}lien Rodriguez, Robert Stojnic, Sergey Edunov,
  and Thomas Scialom.
\newblock Llama 2: Open foundation and fine-tuned chat models.
\newblock {\em CoRR}, abs/2307.09288, 2023.

\bibitem{attention}
Ashish Vaswani, Noam Shazeer, Niki Parmar, Jakob Uszkoreit, Llion Jones, Aidan~N. Gomez, Lukasz Kaiser, and Illia Polosukhin.
\newblock Attention is all you need.
\newblock In Isabelle Guyon, Ulrike von Luxburg, Samy Bengio, Hanna~M. Wallach, Rob Fergus, S.~V.~N. Vishwanathan, and Roman Garnett, editors, {\em Advances in Neural Information Processing Systems 30: Annual Conference on Neural Information Processing Systems 2017, December 4-9, 2017, Long Beach, CA, {USA}}, pages 5998--6008, 2017.

\bibitem{flanv2}
Jason Wei, Maarten Bosma, Vincent~Y. Zhao, Kelvin Guu, Adams~Wei Yu, Brian Lester, Nan Du, Andrew~M. Dai, and Quoc~V. Le.
\newblock Finetuned language models are zero-shot learners.
\newblock {\em CoRR}, abs/2109.01652, 2021.

\bibitem{app3_emergent}
Jason Wei, Yi~Tay, Rishi Bommasani, Colin Raffel, Barret Zoph, Sebastian Borgeaud, Dani Yogatama, Maarten Bosma, Denny Zhou, Donald Metzler, Ed~H. Chi, Tatsunori Hashimoto, Oriol Vinyals, Percy Liang, Jeff Dean, and William Fedus.
\newblock Emergent abilities of large language models.
\newblock {\em Trans. Mach. Learn. Res.}, 2022, 2022.

\bibitem{Outlier}
Xiuying Wei, Yunchen Zhang, Yuhang Li, Xiangguo Zhang, Ruihao Gong, Jinyang Guo, and Xianglong Liu.
\newblock Outlier suppression+: Accurate quantization of large language models by equivalent and effective shifting and scaling.
\newblock In Houda Bouamor, Juan Pino, and Kalika Bali, editors, {\em Proceedings of the 2023 Conference on Empirical Methods in Natural Language Processing, {EMNLP} 2023, Singapore, December 6-10, 2023}, pages 1648--1665. Association for Computational Linguistics, 2023.

\bibitem{wen2024autodroid}
Hao Wen, Yuanchun Li, Guohong Liu, Shanhui Zhao, Tao Yu, Toby Jia-Jun Li, Shiqi Jiang, Yunhao Liu, Yaqin Zhang, and Yunxin Liu.
\newblock Autodroid: Llm-powered task automation in android.
\newblock In {\em Proceedings of the 30th Annual International Conference on Mobile Computing and Networking (MobiCom '24)}, Washington D.C., USA, 2024. Association for Computing Machinery.

\bibitem{application_1}
Carole-Jean Wu, David Brooks, Kevin Chen, Douglas Chen, Sy~Choudhury, Marat Dukhan, Kim Hazelwood, Eldad Isaac, Yangqing Jia, Bill Jia, et~al.
\newblock Machine learning at facebook: Understanding inference at the edge.
\newblock In {\em 2019 IEEE international symposium on high performance computer architecture (HPCA)}, pages 331--344. IEEE, 2019.

\bibitem{wu2024heterogeneity}
Yebo Wu, Li~Li, Chunlin Tian, Tao Chang, Chi Lin, Cong Wang, and Cheng-Zhong Xu.
\newblock Heterogeneity-aware memory efficient federated learning via progressive layer freezing.
\newblock In {\em 2024 IEEE/ACM 32nd International Symposium on Quality of Service (IWQoS)}, pages 1--10. IEEE, 2024.

\bibitem{wu2025survey}
Yebo Wu, Chunlin Tian, Jingguang Li, He~Sun, Kahou Tam, Li~Li, and Chengzhong Xu.
\newblock A survey on federated fine-tuning of large language models.
\newblock {\em arXiv preprint arXiv:2503.12016}, 2025.

\bibitem{yeboSurvey}
Yebo Wu, Chunlin Tian, Jingguang Li, He~Sun, Kahou Tam, Li~Li, and Chengzhong Xu.
\newblock A survey on federated fine-tuning of large language models.
\newblock {\em CoRR}, abs/2503.12016, 2025.

\bibitem{SmoothQuant}
Guangxuan Xiao, Ji~Lin, Micka{\"{e}}l Seznec, Hao Wu, Julien Demouth, and Song Han.
\newblock Smoothquant: Accurate and efficient post-training quantization for large language models.
\newblock In Andreas Krause, Emma Brunskill, Kyunghyun Cho, Barbara Engelhardt, Sivan Sabato, and Jonathan Scarlett, editors, {\em International Conference on Machine Learning, {ICML} 2023, 23-29 July 2023, Honolulu, Hawaii, {USA}}, volume 202 of {\em Proceedings of Machine Learning Research}, pages 38087--38099. {PMLR}, 2023.

\bibitem{yang2023llm4drive}
Zhenjie Yang, Xiaosong Jia, Hongyang Li, and Junchi Yan.
\newblock Llm4drive: A survey of large language models for autonomous driving.
\newblock {\em arXiv preprint arXiv:2311.01043}, 2023.

\bibitem{app6_laptops}
Xinyu Ye, Zhe Wang, Haihao Shen, Yu~Luo, and Hanwen Chang.
\newblock Creating large language models on your laptop.
\newblock \href{https://medium.com/intel-analytics-software/creating-your-own-llms-on-your-laptop-a08cc4f7c91b}{Medium}, 2023.

\bibitem{yi2024phonelm}
Rongjie Yi, Xiang Li, Weikai Xie, Zhenyan Lu, Chenghua Wang, Ao~Zhou, Shangguang Wang, Xiwen Zhang, and Mengwei Xu.
\newblock Phonelm: An efficient and capable small language model family through principled pre-training.
\newblock {\em arXiv preprint arXiv:2411.05046}, 2024.

\bibitem{ELMS}
Wangsong Yin, Rongjie Yi, Daliang Xu, Gang Huang, Mengwei Xu, and Xuanzhe Liu.
\newblock Elms: Elasticized large language models on mobile devices.
\newblock {\em arXiv preprint arXiv:2409.09071}, 2024.

\bibitem{Orca}
Gyeong{-}In Yu, Joo~Seong Jeong, Geon{-}Woo Kim, Soojeong Kim, and Byung{-}Gon Chun.
\newblock Orca: {A} distributed serving system for transformer-based generative models.
\newblock In Marcos~K. Aguilera and Hakim Weatherspoon, editors, {\em 16th {USENIX} Symposium on Operating Systems Design and Implementation, {OSDI} 2022, Carlsbad, CA, USA, July 11-13, 2022}, pages 521--538. {USENIX} Association, 2022.

\bibitem{yuan2023mobile}
Jinliang Yuan, Chen Yang, Dongqi Cai, Shihe Wang, Xin Yuan, Zeling Zhang, Xiang Li, Dingge Zhang, Hanzi Mei, Xianqing Jia, et~al.
\newblock Mobile foundation model as firmware.
\newblock {\em arXiv preprint arXiv:2308.14363}, 2023.

\bibitem{LoRAMoE_1}
Ted Zadouri, Ahmet {\"{U}}st{\"{u}}n, Arash Ahmadian, Beyza Ermis, Acyr Locatelli, and Sara Hooker.
\newblock Pushing mixture of experts to the limit: Extremely parameter efficient moe for instruction tuning.
\newblock {\em CoRR}, abs/2309.05444, 2023.

\bibitem{hellaswag}
Rowan Zellers, Ari Holtzman, Yonatan Bisk, Ali Farhadi, and Yejin Choi.
\newblock Hellaswag: Can a machine really finish your sentence?
\newblock In Anna Korhonen, David~R. Traum, and Llu{\'{\i}}s M{\`{a}}rquez, editors, {\em Proceedings of the 57th Conference of the Association for Computational Linguistics, {ACL} 2019, Florence, Italy, July 28- August 2, 2019, Volume 1: Long Papers}, pages 4791--4800. Association for Computational Linguistics, 2019.

\bibitem{zhan2024heterogeneity}
Shichen Zhan, Yebo Wu, Chunlin Tian, Yan Zhao, and Li~Li.
\newblock Heterogeneity-aware coordination for federated learning via stitching pre-trained blocks.
\newblock In {\em 2024 IEEE/ACM 32nd International Symposium on Quality of Service (IWQoS)}, pages 1--10. IEEE, 2024.

\bibitem{zhao2024felix}
Yifan Zhao, Hashim Sharif, Vikram Adve, and Sasa Misailovic.
\newblock Felix: Optimizing tensor programs with gradient descent.
\newblock In {\em Proceedings of the 29th ACM International Conference on Architectural Support for Programming Languages and Operating Systems, Volume 3}, pages 367--381, 2024.

\bibitem{zhao2023approxcaliper}
Yifan Zhao, Hashim Sharif, Peter Pao-Huang, Vatsin Shah, Arun~Narenthiran Sivakumar, Mateus Valverde~Gasparino, Abdulrahman Mahmoud, Nathan Zhao, Sarita Adve, Girish Chowdhary, et~al.
\newblock Approxcaliper: A programmable framework for application-aware neural network optimization.
\newblock {\em Proceedings of Machine Learning and Systems}, 5, 2023.

\bibitem{zhao7}
Yuanzhe Zhao, Pengyu He, Yan Zhu, Rui~P Martins, Chi-Hang Chan, and Minglei Zhang.
\newblock A 28-nm 3.32-nj/frame compute-in-memory cnn processor with layer fusion for always-on applications.
\newblock {\em IEEE Transactions on Circuits and Systems I: Regular Papers}, 2025.

\bibitem{zhao5}
Yuanzhe Zhao, Yang Wang, Yuheng Wang, Heng Xie, Yan Zhu, RP~Martins, Chi-Hang Chan, Shouyi Yin, and Minglei Zhang.
\newblock A 28nm value-wise hybrid-domain compute-in-memory macro with heterogeneous memory fabric and asynchronous sparsity manager.
\newblock In {\em 2025 IEEE Custom Integrated Circuits Conference (CICC)}, pages 1--3. IEEE, 2025.

\bibitem{zhao4}
Yuanzhe Zhao, Yuheng Wang, Zijian Wang, Yan Zhu, RP~Martins, Chi-Hang Chan, and Minglei Zhang.
\newblock A reconfigurable 0.69-1.02 nj/classification biomedical ai processor for intelligent health monitoring devices.
\newblock In {\em 2025 IEEE Custom Integrated Circuits Conference (CICC)}, pages 1--3. IEEE, 2025.

\bibitem{zhao6}
Yuanzhe Zhao, Heng Xie, Zijian Wang, Chunlin Tian, Li~Li, Yan Zhu, RP~Martins, Chi-Hang Chan, and Minglei Zhang.
\newblock A one-shot floating-point compute-in-memory macro featuring pvt robustness and mismatch tolerance for edge llms.
\newblock In {\em 2025 IEEE Custom Integrated Circuits Conference (CICC)}, pages 1--3. IEEE, 2025.

\bibitem{zhao3}
Yuanzhe Zhao, Minglei Zhang, Pengyu He, Yan Zhu, Chi{-}Hang Chan, and Rui~Paulo Martins.
\newblock A double-mode sparse compute-in-memory macro with reconfigurable single and dual layer computation.
\newblock In {\em {IEEE} Custom Integrated Circuits Conference, {CICC} 2023, San Antonio, TX, USA, April 23-26, 2023}, pages 1--2. {IEEE}, 2023.

\bibitem{vicuna}
Lianmin Zheng, Wei{-}Lin Chiang, Ying Sheng, Siyuan Zhuang, Zhanghao Wu, Yonghao Zhuang, Zi~Lin, Zhuohan Li, Dacheng Li, Eric~P. Xing, Hao Zhang, Joseph~E. Gonzalez, and Ion Stoica.
\newblock Judging llm-as-a-judge with mt-bench and chatbot arena.
\newblock In Alice Oh, Tristan Naumann, Amir Globerson, Kate Saenko, Moritz Hardt, and Sergey Levine, editors, {\em Advances in Neural Information Processing Systems 36: Annual Conference on Neural Information Processing Systems 2023, NeurIPS 2023, New Orleans, LA, USA, December 10 - 16, 2023}, 2023.

\bibitem{Event-based}
Yuhao Zhu, Matthew Halpern, and Vijay~Janapa Reddi.
\newblock Event-based scheduling for energy-efficient qos (eqos) in mobile web applications.
\newblock In {\em 21st {IEEE} International Symposium on High Performance Computer Architecture, {HPCA} 2015, Burlingame, CA, USA, February 7-11, 2015}, pages 137--149. {IEEE} Computer Society, 2015.

\bibitem{bilgili2024llmrobot}
Ömer Bilgin~Bilgili.
\newblock Ai robotics case - controlling robots with llms (large language models).
\newblock \href{https://acrome.net/post/controlling-robots-with-llms-large-language-models}{acrome}, 2024.
\newblock Acrome Robotics Blog, October 23, 2024.

\end{thebibliography}


\end{document}